\documentclass[fleqn]{2023SCGE}
\usepackage{aas_macros}
\usepackage{amssymb,amsmath}
\usepackage{xspace}
\usepackage{multirow,array}
\usepackage[percent]{overpic}

\graphicspath{{./}{figures/}{figures/extra/}}

\newcommand{\msunh}{h^{-1}\mathrm{M}_\odot}
\newcommand{\mpch}{h^{-1}{\rm Mpc}}
\newcommand{\kpch}{h^{-1}{\rm kpc}}
\newcommand{\gpch}{h^{-1}{\rm Gpc}}

\newcommand{\hbt}{\textsc{hbt+}\xspace}
\newcommand{\fof}{\textsc{FoF}\xspace}
\newcommand{\LCDM}{$\Lambda$CDM\xspace}

\newcommand{\lgadget}{\textsc{LGadget}\xspace}
\newcommand{\gadget}{\textsc{Gadget}\xspace}
\newcommand{\subfind}{\textsc{SubFind}\xspace}
\newcommand{\ltree}{\textsc{LHaloTrees}\xspace}
\newcommand{\lgalaxy}{\textsc{LGalaxies}\xspace}
\newcommand{\lgalaxyp}{\textsc{LGalaxies-Line}\xspace}
\newcommand{\lgalaxyl}{\textsc{LGalaxies-Lensing}\xspace}
\newcommand{\gaea}{\textsc{Gaea}\xspace}
\newcommand{\blic}{\textsc{Blic}\xspace}
\newcommand{\starduster}{\textsc{StarDuster}\xspace}
\newcommand{\cube}{\textsc{Cube}\xspace}

\newcommand{\rev}[1]{#1} 

\begin{document}

\ensubject{subject}

\ArticleType{Article}
\SpecialTopic{SPECIAL TOPIC: }
\Year{2023}
\Month{January}
\Vol{66}
\No{1}
\DOI{??}
\ArtNo{000000}
\ReceiveDate{--}
\AcceptDate{--}

\title{The Jiutian simulations for the CSST extra-galactic surveys}{The Jiutian simulations for the CSST extra-galactic surveys}


\author[1,2]{Jiaxin Han}{{jiaxin.han@sjtu.edu.cn}}%
\author[3]{Ming Li}{}
\author[1,2]{Wenkang Jiang}{}
\author[1,2]{Zhao Chen}{}
\author[4,5]{Huiyuan Wang}{}
\author[6]{Chengliang Wei}{}
\author[1,2]{Feihong He}{}
\author[7]{\\Jian-hua He}{}
\author[8]{Jiajun Zhang}{}
\author[9]{Yu Liu}{}
\author[10,11,12]{Weiguang Cui}{}
\author[1,2]{Yizhou Gu}{}
\author[3]{Qi Guo}{}
\author[1,2]{Yipeng Jing}{ypjing@sjtu.edu.cn}
\author[13]{Xi Kang}{}
\author[6]{\\Guoliang Li}{}
\author[4,5]{Xiong Luo}{}
\author[14]{Yu Luo}{}
\author[15]{Wenxiang Pei}{}
\author[16]{Yisheng Qiu}{}
\author[1,2]{Zhenlin Tan}{}
\author[17]{Lizhi Xie}{}
\author[1,2]{\\Xiaohu Yang}{}
\author[18]{Hao-Ran Yu}{}
\author[1,2]{Yu Yu}{}
\author[1,2]{Jiale Zhou}{}

\address[1]{Department of Astronomy, School of Physics and Astronomy, Shanghai Jiao Tong University, Shanghai, 200240, China}
\address[2]{State Key Laboratory of Dark Matter Physics, Key Laboratory for Particle Astrophysics and Cosmology (MOE), \\\&Shanghai Key Laboratory for Particle Physics and Cosmology, Shanghai Jiao Tong University, Shanghai, 200240, China}
\address[3]{Key Laboratory for Computational Astrophysics, National Astronomical Observatories, Chinese Academy of Sciences, Beijing 100101, China}
\address[4]{Department of Astronomy, University of Science and Technology of China, Hefei, Anhui 230026, China}
\address[5]{School of Astronomy and Space Science, University of Science and Technology of China, Hefei 230026, China}
\address[6]{Purple Mountain Observatory, Nanjing 210008, China}
\address[7]{School of Astronomy and Space Science, Nanjing University, Nanjing 210093, P. R. China }
\address[8]{Shanghai Astronomical Observatory, Chinese Academy of Sciences, Shanghai 200030, China}
\address[9]{Department of Astronomy, Tsinghua University, Beijing, 100084, P.R. China}
\address[10]{Departamento de Física Teórica, M-8, Universidad Autónoma de Madrid, Cantoblanco 28049, Madrid, Spain}
\address[11]{Centro de Investigaci\'{o}n Avanzada en F\'{i}sica Fundamental (CIAFF), Facultad de Ciencias, Universidad Aut\'{o}noma de Madrid, E-28049, Madrid, Spain}
\address[12]{Institute for Astronomy, University of Edinburgh, Royal Observatory, Edinburgh EH9 3HJ, United Kingdom}
\address[13]{Institute for Astronomy, School of Physics, Zhejiang University, Hangzhou 310027, China}
\address[14]{Department of Physics, School of Physics and Electronics, Hunan Normal University, Changsha 410081, China}
\address[15]{Shanghai Key Lab for Astrophysics, Shanghai Normal University, Shanghai 200234, People’s Republic of China}
\address[16]{Research Center for Astronomical Computing, Zhejiang Laboratory, Hangzhou, China}
\address[17]{Astrophysics Center, Tianjin Normal University, Tianjin, 300387, China}
\address[18]{Department of Astronomy, Xiamen University, Xiamen, Fujian 361005, China}

\AuthorMark{Han J., et al.}

\AuthorCitation{Han J., et al}

\abstract{We provide an overview of the Jiutian simulations, a hybrid simulation suite for the China Space Survey Telescope (CSST) extragalactic surveys. It consists of four complementary modules: the primary runs with high resolutions \rev{with} the fiducial concordance cosmology, the emulator runs exploring the parameter uncertainties around the fiducial cosmology, the reconstruction runs intended for recovering the observed Universe \rev{position by position}, and the extension runs employing extended cosmologies beyond the standard model. For the primary runs, two independent pipelines are adopted to construct subhaloes and merger trees. On top of them, four sets of mock galaxy light-cone catalogs are produced from semi-analytical models and subhalo abundance matching, providing a variety of observational properties including galaxy SED, emission lines, lensing distortions, and mock images. The 129 emulator runs are used to train the CSST emulator, achieving one percent accuracy in predicting \rev{the matter power spectrum} over $k\leq 10 h{\rm Mpc}^{-1}$ and $z\leq 2$. The reconstruction runs employ a number of subgrid baryonic models to predict the evolution and galaxy population \rev{resembling} certain regions in the real Universe with constrained initial conditions, enabling controlled investigation of galaxy formation on top of structure formation. The extension runs cover models with warm dark matter, $f(R)$ gravity, interacting dark energy, and nonzero neutrino masses, revealing differences in the cosmic structure under alternative cosmological models. We introduce the specifications for each run, the data products derived from them, the corresponding pipeline developments, and present some main tests. Using the primary runs, we also show that the subhalo peak mass functions of different levels are approximately universal. These simulations form a comprehensive and open simulation library for CSST surveys and beyond.}


\keywords{cosmology, large scale structure of the Universe, computer modeling and simulation, astronomical catalogs}

\PACS{98.80.-k, 98.65.-r, 95.75.Pq, 95.80.+p}

\maketitle

\begin{multicols}{2}
\section{Introduction} \label{sec:intro}
Cosmological simulations are nowadays an essential engine powering the design and interpretation of modern galaxy surveys. The simulated galaxy sample prior to observational effects can serve as the starting point for assessing the effectiveness of the survey strategy, testing various instrumental systematics, and developing and validating the survey analysis pipelines. Mock galaxy catalogs built on top of the simulations with added observational effects enable end-to-end comparisons between theory and observations, which help to test our understandings for cosmology, structure formation and galaxy formation. 

With the increase in computational power and developments of numerical algorithms, cosmological simulations are constantly improving both in the spatial coverage and mass resolution, parallelizing the advancement in the survey scales. For example, the Millennium simulation~\cite{Millennium} carried out two decades ago covers a boxsize of 500 $\mpch$ using $2160^3$ particles, corresponding to a mass resolution of $m_p=8.6\times10^8\msunh$. Nowadays one of the largest simulations, the Uchuu simulation~\cite{Uchuu}, boosted the boxsize to 2000 $\mpch$ using $12800^3$ particles, resulting in a similar particle mass of $3.27\times 10^8\msunh$, while the Shin-Uchuu run enhanced the mass resolution to $8.97\times 10^5\msunh$ using $6400^3$ particles in a box of size $140\mpch$. To help with model exploration and parameter inference, it has also become common to carry out multiple simulation runs under different cosmologies, or different realizations of the same cosmology. For example, the AbacusSummit simulations~\cite{AbacusRun,AbacusCode}, designed to meet the simulation requirements of the Dark Energy Spectroscopic Instrument (DESI) survey~\cite{DESI}, come with $\sim$ 2000 small runs each of comparable specifications to the Millennium run, in addition to a small number of much larger runs with bigger box sizes or better mass resolutions. The small runs are useful for estimating covariances of cosmological statistics, while the large runs serve to provide detailed predictions of the cosmic structure. On the other hand, the DarkQuest\cite{DarkQuest} and Aemulus\cite{Aemulus,AemulusIII} projects employ $\sim100$ simulations with varying cosmological parameters to train cosmological emulators for fast predictions of major statistics of the large scale structure.

While a simulation is not necessarily tied up to a given
survey, the ever expanding size of the simulation data can make
it intractable to keep or share the full simulation snapshots for complete re-analysis under other scientific requirements, due to limitations in storage, network and processor resources. It is also \rev{beneficial} to have competing simulations run using different simulation codes and post-processing pipelines to enable cross-validations, as well as to drive their developments. As a result, even though a number of large simulations have been released publicly, it is still advantageous for different communities to have their own simulation repositories. For example, analogous to the connection between AbacusSummit and DESI, the EuclidFlagship~\cite{PKDGRAV3,EuclidFlagship} simulations are carried out within the Euclid~\cite{EUCLID} mission.

In this work, we introduce Jiutian,\footnote{Jiutian refers to the nine (or many) realms of the Universe in ancient Chinese.} a hybrid suite of cosmological simulations intended for supporting the scientific analysis of the China Space Survey Telescope (CSST) extra-galactic surveys. The CSST is a 2 meter space telescope to be launched next to the Tian Gong space station. It is designed to have a field of view of at least 1.7 square degree and an angular resolution better than 0.13 arc-seconds. The telescope will be equipped with multiple detector instruments serving different observational purposes. Among them, the survey module is composed of a primary focus camera capable of carrying out multi-band imaging and slitless spectroscopy simultaneously. Over a 10-year timescale, the telescope will survey around 40\% of the sky from NUV to the infrared, providing a comprehensive database for studying stellar physics, galaxy formation and cosmology\cite{2011SSPMA..41.1441Z, 2018MNRAS.480.2178C, 2019ApJ...883..203G, ZhanHu2021...CSB}.

\begin{table*}[t]
 \footnotesize
\begin{threeparttable}
\caption{Jiutian Simulation Modules}
\label{tab:jiutian_overview}
\tabcolsep 18pt 
\begin{tabular*}{\textwidth}{ccccc}
\toprule
\hline
         &  Cosmology & Type &  Resolution & Number of runs\tnote{1)} \\\hline
         Primary & $\Lambda$CDM, Planck 2018 & $N$-body & high & 3\\\hline
 Emulator& $w$CDM + neutrino\tnote{2)} & $N$-body  & medium & 128+1 \\\hline
  Reconstruction& constrained $\Lambda$CDM & zoom-in hydrodynamical & high & 9\\\hline
 \multirow{4}{*}{Extension}&  WDM  & $N$-body & high & 3+1 \\\cline{2-5}
 & $f(R)$ gravity & \multirow{3}{*}{$N$-body} & \multirow{3}{*}{low} & 3+1 \\
 & Interacting Dark Energy & & & 6+1\\
 & $\Lambda$CDM + neutrino & & & 4+2\\\hline
 \bottomrule
 \end{tabular*}
 \begin{tablenotes}
     \item[1)] The number after the plus sign indicates the number of matching \LCDM runs in each extension model.
     \item[2)] CDM cosmology with a redshift dependent equation of state for dark energy and the existence of massive neutrinos.
 \end{tablenotes}
\end{threeparttable}
\end{table*}

The Jiutian simulations are composed of four complementary modules as follows.
\begin{itemize}
    \item \textbf{Primary}: high resolution runs of the concordance cosmological model, with varying boxsizes and particle masses. These runs target the resolution and volume requirements of the CSST survey.
    \item \textbf{Emulator}: a large number of medium resolution runs covering the parameter uncertainties of the (generalized) concordance cosmological model. These runs are used for training the CSST emulator.
    \item \textbf{Reconstruction}: hydrodynamical zoom-in simulations of certain regions of the observed Universe, according to the reconstructed initial conditions from SDSS observations.
    \item \textbf{Extension}: low to high resolution runs exploring extensions or modifications to the standard cosmological model, including $\Lambda$CDM with different neutrino masses, $f(R)$ gravity model and interacting dark energy model.
\end{itemize}
The specifications of these simulations are summarized in Table~\ref{tab:jiutian_overview}. These four modules complement each other by focusing on the resolution, parameter uncertainty, model space and realism of the simulated universes, respectively.

The Jiutian simulations have already found applications in a number of studies on cosmology and galaxy formation as well as in CSST pipeline developments~\cite{app1,app2,app3,app4,app5,app6,app7,app8,app9,app10,app11,Shi25BAO,Ma25}. The purpose of this paper is thus to provide an overview of the simulations and the data products, and to serve as a standard reference for using these simulations.

In the following sections we introduce the four modules in detail, including their specifications, main results and data products. We summarize and conclude in Section~\ref{sec:summary}.

\section{Primary runs}\label{primary}
The primary runs are the flagship runs in the Jiutian suite, aiming to meet the depth and volume requirement of the CSST extra-galactic surveys. 

\subsection{CSST survey requirement}
The CSST imaging surveys consist of a deep survey and an ultra deep survey. The nominal flux limits are as follows:
\begin{itemize}
\item Deep: $g<26$, $y<24.4$, $u,r,i,z,NUV<25.5$,
\item Ultra deep: $g<27$, $y<25.7$, $u,r,i,z,NUV<26.5$.
\end{itemize}
The spectroscopic surveys will cover the same footprints as the imaging surveys, but to \rev{brighter} flux limits. 

To investigate the simulation requirements of these surveys, we first construct a preliminary mock catalog from existing semi-analytical galaxy catalogs in the Millennium database.\footnote{\url{https://gavo.mpa-garching.mpg.de/MyMillennium/}} We select galaxies according to the flux limits\footnote{As the database does not provide magnitudes in the CSST filters, we approximate them with $u,g,r,i,z$ magnitudes in SDSS, $NUV$ in GALEX and $Y$ in VISTA filters.} of the surveys from the \textsc{Henriques15-MRscPlanck1-M05}~\cite{Henriques15} light-cones in the database.  Fig.~\ref{fig:millennium_distr} shows the distribution in halo mass and redshift of the selected galaxies. For a conservative estimation, we have aggregated all the galaxies that pass the magnitude selection in at least one band. 

As shown in the left panel, for $z>1$, the majority of galaxies in the deep survey reside in halos with a peak mass above $10^{11}\msunh$. Resolving these halos with at least 100 particles requires a particle mass of $10^9\msunh$. For the ultra deep survey, the mass resolution requirement is increased by a factor of $\sim 3$.  The right panel of Fig.~\ref{fig:millennium_distr} shows the cumulative redshift distribution of these galaxies. The majority of these galaxies are distributed in the redshift range of about $1-3$. If we apply a more stringent selection by only keeping galaxies detectable in all the CSST bands, the maximum redshift shrinks to $z\sim 1.5$, corresponding to a comoving distance of $\sim 3\gpch$. This much lower redshift limit is mostly caused by the Lyman break redshifted to the NUV band at $z\sim 1-2$, resulting in a significant reduction in the NUV magnitude.



\subsection{Simulations and particle lightcones}
\begin{figure}[H]
\centering
\includegraphics[width=0.5\textwidth]{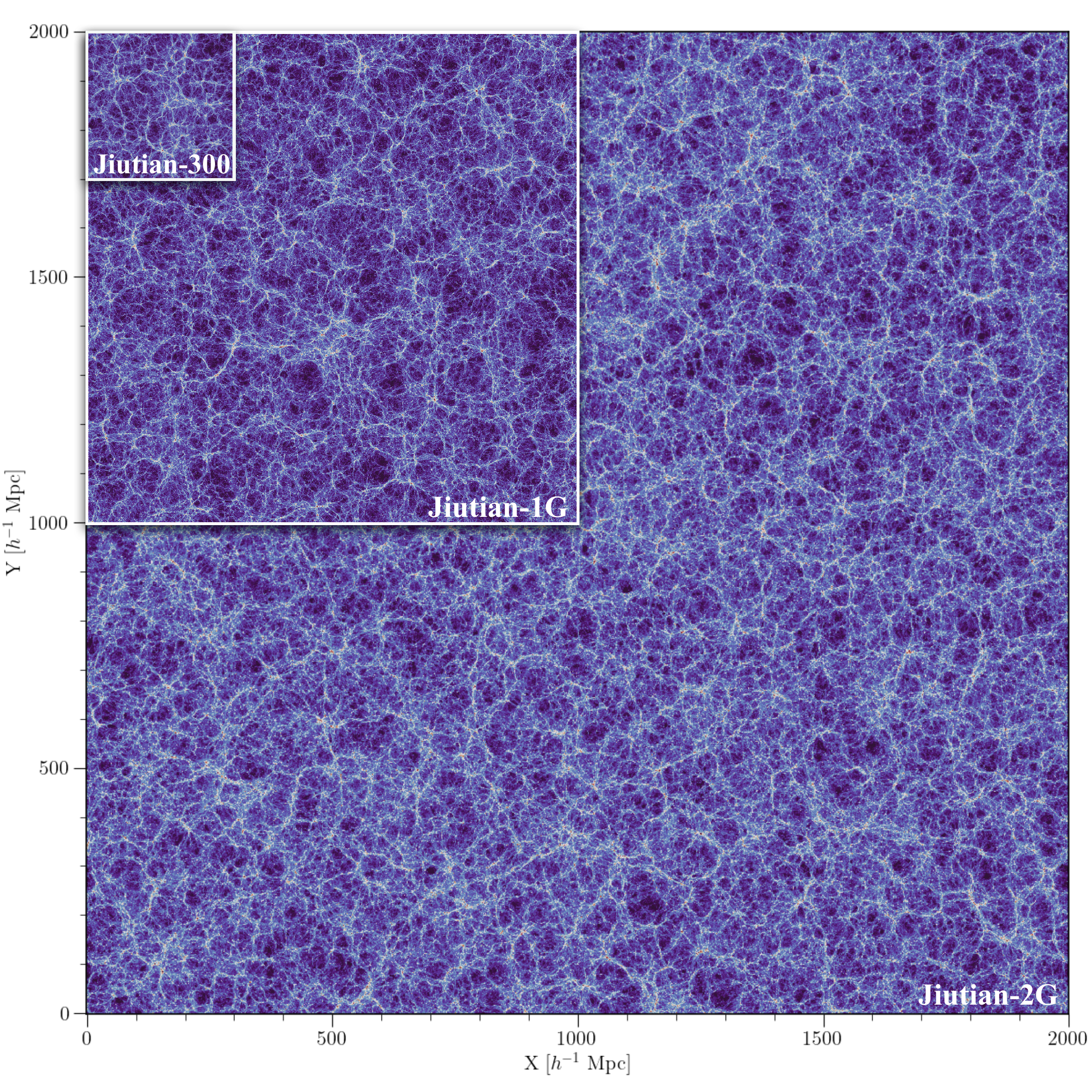}
\caption{Projected density maps of the Jiutian primary runs, overlaid and aligned at the top-left corner. The projection is over a thickness $30\ \mpch$, and each image is resolved in $2000^2$ pixels. The three simulations follow the same cosmology with an identical number of particles, except for different box sizes.}
\label{fig:JT-Prim-fig}
\end{figure}

Targeting the CSST survey requirements with our available computational resources, three dark matter only simulations are conducted using $6144^3$ particles with different boxsizes, as detailed in Table~\ref{tab:primary}. They adopt the $\Lambda$ Cold Dark Matter (\LCDM) cosmological model with parameters according to the Planck-2018~(\cite{Planck2018}; last column of their Table 2) results, with $\Omega_{\rm m}=0.3111$, $\Omega_\Lambda=0.6889$, $\Omega_{\rm b}=0.049$, $n_s=0.9665$, $\sigma_8=0.8102$ and $H_0=0.6766$.  Figure~\ref{fig:JT-Prim-fig} shows a visualization of the matter distribution in the three runs.

\begin{figure*}
\includegraphics[width=\textwidth]{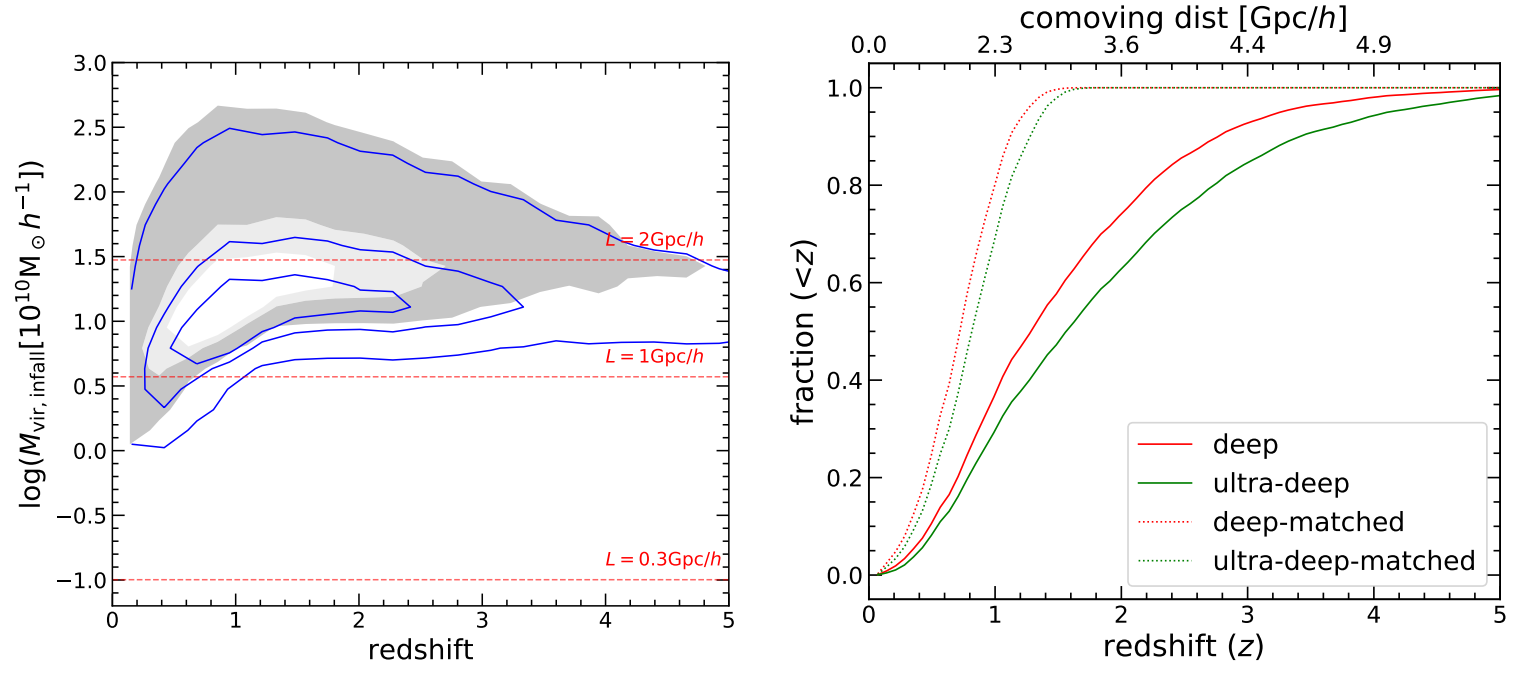}
\caption{\textit{Left}: Distribution of CSST galaxies in halo mass and redshift, aggregated from six bands. The galaxies are selected from the Millennium light-cone catalog if they can pass any one of the CSST flux cuts. The contours show the most-crowded regions containing 30\%, 60\% and 95\% of observed galaxies. The results for the deep and ultra-deep surveys are shown as gray filled and blue unfilled contours, respectively. The halo mass corresponding to 100 particles in the three primary Jiutian runs are shown as red dashed lines. \textit{Right}: The cumulative fraction of CSST galaxies over redshift (bottom axis) and comoving distance (top axis). The solid curves show the distribution of galaxies aggregated from the six bands, while the dotted curves show the distributions of galaxies jointly observable across the bands. The red and green curves show the results for the deep and ultra deep surveys, respectively.}\label{fig:millennium_distr}
\end{figure*}

The halo mass resolutions at $100$ particles are compared with the CSST survey requirement in the left panel of Fig.~\ref{fig:millennium_distr}. The highest resolution run, Jiutian-300, provides sufficient resolution for resolving nearby dwarf galaxies down to $10^9\msunh$ in halo mass. For $z>1$, Jiutian-1G is sufficient for resolving the host halos for the majority ($>95\%$) of galaxies. Jiutian-2G complements the two by providing a much larger boxsize, useful for providing a large sample of the most massive galaxies as well as for assessing cosmic variances in large-scale statistics. 

\begin{table*}[t]
\centering
\begin{threeparttable}
 \footnotesize
\caption{Specifications of the primary runs. Boxsize and softening are in comoving units.}
\label{tab:primary}
\begin{tabular}{ccccccc}
\toprule
\hline
         &  Boxsize ($\mpch$) & Softening ($\kpch$) &  Particle Mass ($\msunh$) & Particles &  Snapshots & Cosmology \\\hline
         Jiutian-300 & 300 & 1.0 & 1.005$\times 10^7$  & \multirow{3}{*}{$6144^3$}  & \multirow{3}{*}{128} & \multirow{3}{*}{Planck 2018}  \\
 Jiutian-1G & 1000 & 4.0 & 3.723 $\times 10^8$ & &  & \\
 Jiutian-2G & 2000 & 7.0 & 2.978$\times 10^9$ & &  & \\\hline
 \bottomrule
 \end{tabular}
\end{threeparttable}
\end{table*}

Among the three runs, Jiutian-1G was run with \lgadget-3~\cite{Angulo12}, while Jiutian-300 and Jiutian-2G were run with the newly released \gadget-4 code~\cite{Gadget4}. To generate the initial conditions, the linear power spectrum is computed with {\tt CAMB} \footnote{\url{https://camb.info}} for an initial redshift of $z_{\rm ini}=127$. The Zel'dovich approximation is then employed to create the displacement field on $7680^3$ grids for computing the position and velocity perturbations of the unperturbed particle load with a glass-like pre-initial condition. 
\rev{We save 10 snapshots at $z=(127,80, 70, 60, 50, 40, 35, 30, 25,20)$, and 118 snapshots between $z=20$ to $z=0$ according to the spacing scheme
used in the Millennium II simulation~\cite{MillenniumII} (Eq.2 of the reference), but with a doubled number of snapshots.}

\begin{figure}[H]
    \includegraphics[width=0.5\textwidth]{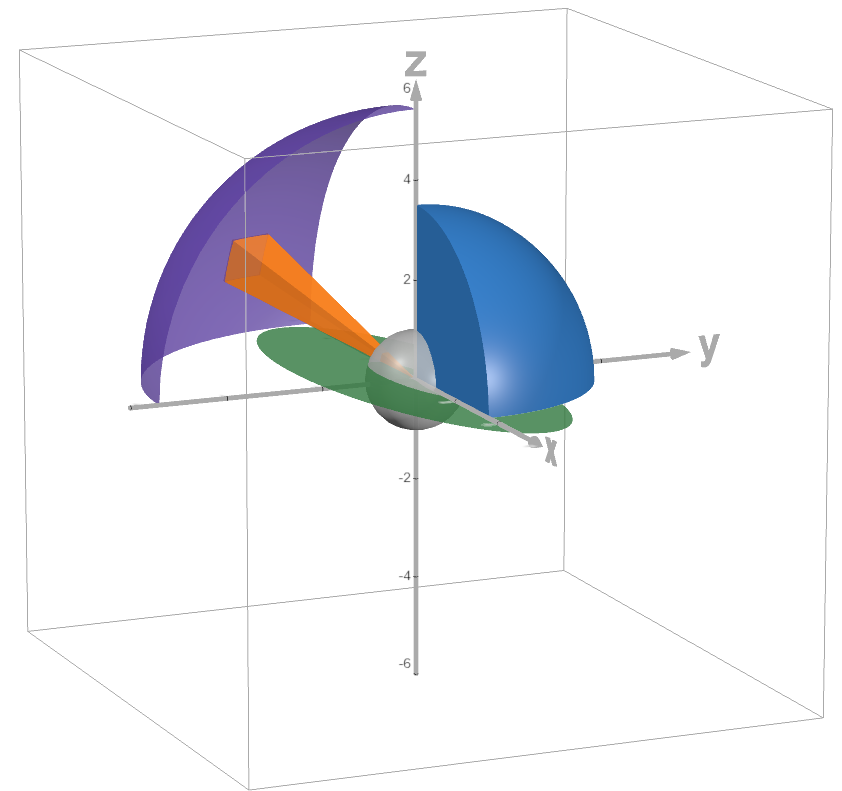}
    \caption{Illustration of the geometries of the five  particle lightcones in Jiutian 2G. From inside out, the five lightcones are LC0 (grey), LC2 (blue), LC3 (green), LC4 (orange) and LC1 (purple). Note LC1 is also a full sky lightcone of most-bound particles while only one quadrant of it is shown for clarity. The coordinates shown are in comoving units of $\gpch$.
    }\label{fig:lightcones}
\end{figure}

For Jiutian-300 and Jiutian-2G, particle lightcones are created on-the-fly with the new feature of \gadget-4. Sticking on the default setting, the fiducial observer is placed at the origin of the simulation \rev{box}, and the positions and velocities of particles are recorded as they cross the past lightcone of the observer.\footnote{\rev{The placement of the observer for these lightcones are not optimized to match the local environment around the Milky Way, which should be taken into account for applications sensitive to the local environment.}} To facilitate different applications, we have created five lightcones of different geometries (as illustrated in Figure~\ref{fig:lightcones}) and contents in Jiutian-2G. 
\begin{itemize}
    \item \textbf{Jiutian-2G-LC0}: a full-sky particle lightcone with a redshift range from 0 to 0.389,\rev{corresponding to a comoving radius of $1\ \gpch$, which is free from duplicated structures arising from the periodic boundary condition};
    \item \textbf{Jiutian-2G-LC1}: a full-sky lightcone only for the most-bound particles of \subfind subhalos,\footnote{A most-bound particle can still be included even after its subhalo becomes unresolved, serving as a fiducial tracer for the subhalo and useful for seeding an orphan galaxy in semi-analytical models of galaxy formation.} with a redshift coverage from 0 to 5, or $\sim 5400 \mpch$;
    \item \textbf{Jiutian-2G-LC2}: an octant-sky particle lightcone ($x>0, y>0, z>0$), extending to $z\sim1.9$, or $\sim 3500 \mpch$; 
    \item \textbf{Jiutian-2G-LC3}: a disk-like lightcone, with a thickness of $15\ h^{-1} {\rm Mpc}$, a disk radius extending to $z=1.9$ ($\sim 3500 \mpch$), and a normal vector of the disk pointing towards the direction of ${\vec{e}_n} =(1, 3, 10)$; 
    \item \textbf{Jiutian-2G-LC4}: a square pencil beam particle lightcone extending to $z=5$ ($\sim 5400 \mpch$), with a field of view of $(10 {\rm deg})^2$ and pointing towards the direction of $(-3, -1, 1)$. This corresponds to a field with an area of $(\sim 940\ \mpch)^2$ at $z=5$. \rev{The direction of the pencil beam is chosen to avoid duplication of particles as much as possible given the boxsize, following \cite{Hilbert09}}.
\end{itemize}

Due to the limited volume coverage and high mass resolution, only three lightcones are produced in Jiutian-300.
\begin{itemize}
    \item \textbf{Jiutian-300-LC0}: a full-sky lightcone only for the most-bound particles of \subfind subhalos, with a redshift coverage from 0 to 0.3 ($\sim 830 \mpch$);
    \item \textbf{Jiutian-300-LC1}: a disk-like lightcone, with a thickness of $15\ h^{-1} {\rm Mpc}$, a disk radius extending to $z=0.25$ ($\sim 700 \mpch$), and a normal vector pointing towards ${\vec{e}_n} =(1, 3, 10)$, ; 
    \item \textbf{Jiutian-300-LC2}: a square pencil beam particle lightcone extending to $z=0.25$ ($\sim 700 \mpch$), with a field of view of $(10 {\rm deg})^2$ and pointing towards $(-3, -1, 1)$.
\end{itemize}

\begin{figure*}
\includegraphics[width=\textwidth]{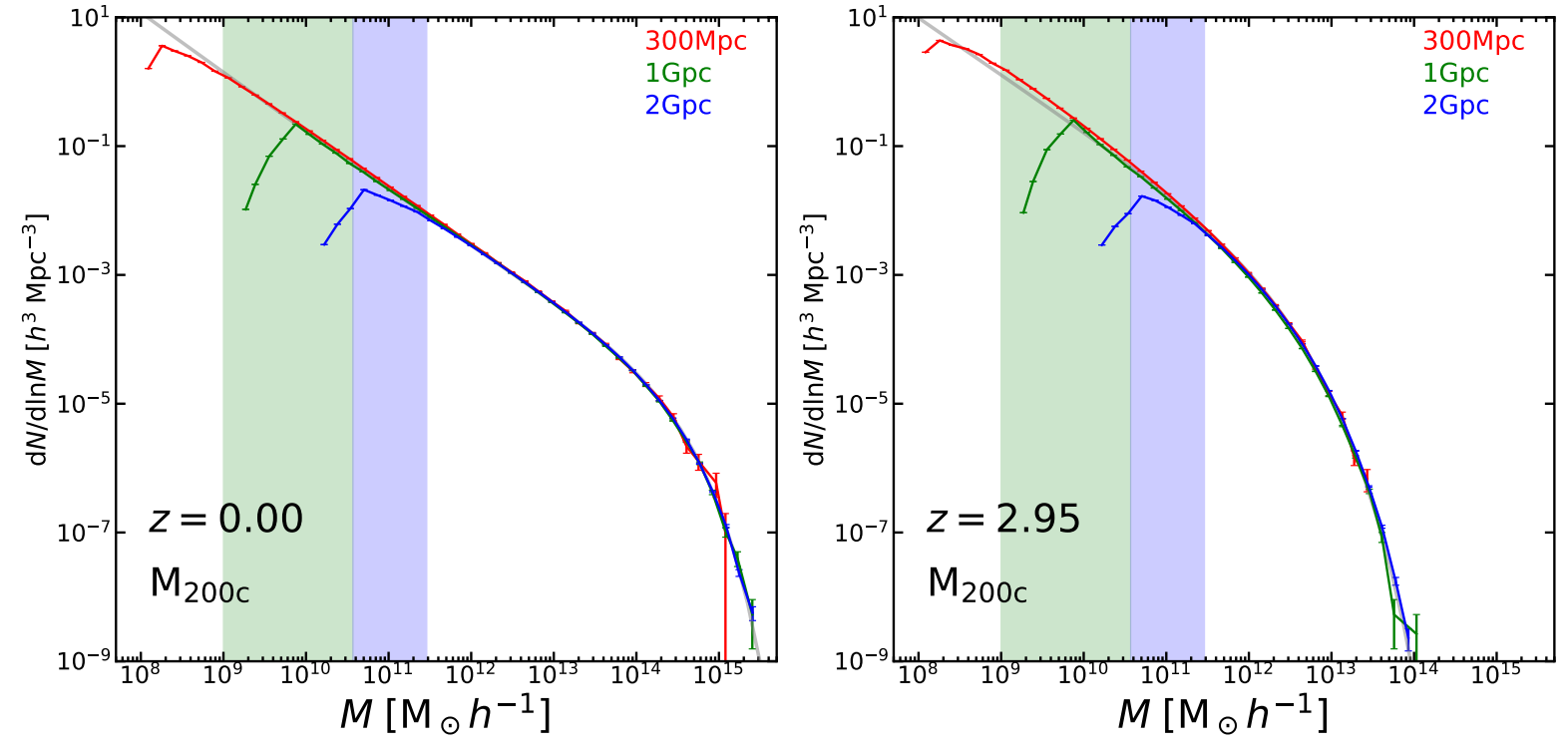}
\caption{Halo mass functions of the Jiutian primary runs at $z=0$ (left) and $z=2.95$ (right). The boundaries of the three shaded regions mark the 100-particle mass limits of the three simulations. In each panel, the solid gray line correspond to fitted halo mass functions from \cite{Watson13} at the corresponding redshift. The error bars represent the Poisson errors in the halo counts. Only self-bound halos containing at least 20 bound particles are included in the measurements. }\label{fig:primaryhmf}
\end{figure*}

\subsection{(Sub)Halo catalogs and merger trees}
Friends-of-friends~(\fof; \cite{FoF}) halos are identified at each snapshot using a linking-length of $0.2$ times the mean inter-particle separation.\footnote{The linking-length used in the \gadget-4 runs is slightly larger due to a different convention in the initial code release, by a factor of $(1/(1-f_b))^{1/3}$, where $f_b$ is the cosmic baryon fraction. This results in $\sim 3\%$ larger $M_{\rm 200c}$ at a fixed abundance or an $\sim 1\%$ overestimation in the halo mass function at $z=0$. The corresponding numbers are $\sim 8\%$ and $\sim 15\%$ at $z=3$, due to the steeper slope of the halo mass function at a higher redshift.
} Fig.~\ref{fig:primaryhmf} shows the halo mass functions of the primary runs, counting only halos that are self-bound. Comparing the three runs, a good convergence is observed for halos resolved with 100 or more particles. The mass functions also agree well with the model from Watson et al. 2013~\cite{Watson13}, who fitted the mass functions of bound halos over a range of cosmologies.

Subhaloes and merger trees are constructed using two independent pipelines. The first pipeline identifies subhaloes at each snapshot with \subfind~\cite{subfind}, a configuration space subhalo finder that identifies subhaloes as self-bound and locally overdense substructures in the host FoF halo. Halo and subhalo merger trees are subsequently built with \ltree~\cite{Millennium, Angulo12, Angulo14} by linking subhaloes across snapshots. The other pipeline adopts \hbt~\cite{hbt,hbtplus}, a time-domain subhalo finder and merger tree builder. It works by tracking the evolution of every halo in the simulation from the earliest to the lastest snapshots. A halo turns into a subhalo tracked by its self-bound particles once it merges into another larger halo. A unique track ID is created to identify each halo throughout its evolution history including the subhalo phase, allowing for easy retrieval of the evolution history. This algorithm naturally avoids many difficulties associated with conventional subhalo finders and is capable of producing highly physical and consistent subhalo catalogs along with their merger histories. To process the Jiutian simulations, \hbt have been updated with an interface to the \gadget-4 output,\footnote{\url{https://github.com/Kambrian/HBTplus/tree/MPI-Hydro-G4}} and the input routines have been restructured to speed up the reading of the halo catalogs.

\begin{figure*}
\includegraphics[width=\textwidth]{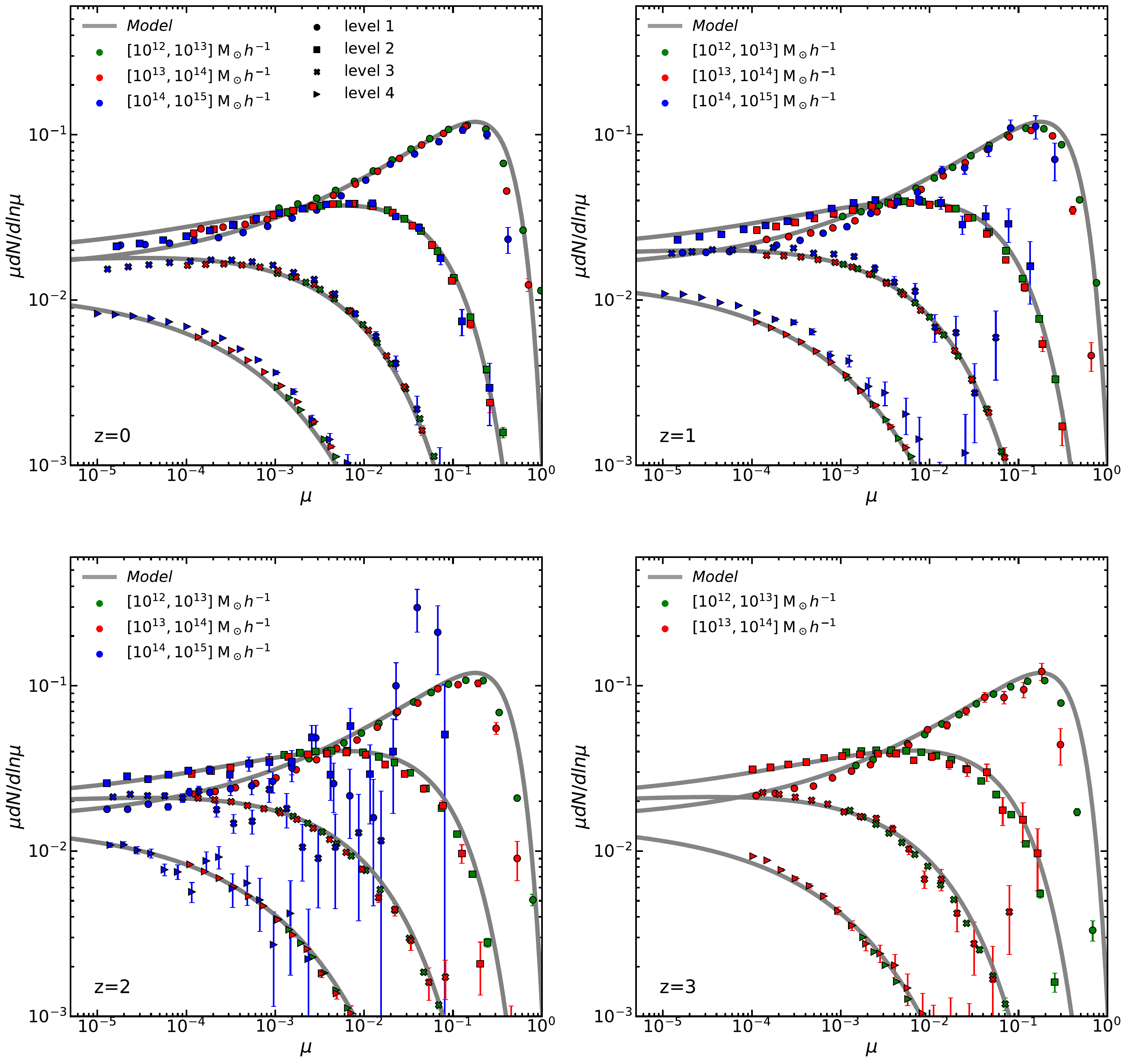}%
\caption{The subhalo peak mass functions (PMFs) of different levels resolved with \hbt in Jiutian-300. A level-1 subhalo is one that infalled directly to the current host, while high level ones are those that fall into lower-level hosts first before being merged into the current host. The PMFs are measured for different host halo mass bins as indicated by the different colors, while the results for different levels are shown with different symbols, as labeled. The error bars indicate Poisson noise in the counts. The gray solid curves represent fitting results using the analytical PMF model of Jiang et al. (2025)\cite{Jiang25}. The model fits the lelvel-1 PMFs with a universal function, and derives the high-level PMFs as self-convolutions of the level-1 PMF (see text for details).}\label{fig:pmf_levels}
\end{figure*}

The use of \hbt also allows for the decomposition of the subhalo population into different merger levels. The subhalo level is recorded with the \texttt{Depth} variable in the subhalo catalog of \hbt. A level-1 subhalo is directly accreted into the host halo, while a level-2 subhalo is accreted into a level-1 subhalo first before they merge into the current host halo together. The level distribution of subhaloes thus encodes the hierarchical merger histories of their progenitor halos, and remains one of the relatively less studied properties of subhaloes. To illustrate the potential of the Jiutian simulations in studying the subhaloes at different levels, in Fig.~\ref{fig:pmf_levels} we show the subhalo peak mass functions (PMF) measured in the Jiutian-300 simulation at a few redshifts, in which the peak mass of a subhalo is defined as the maximum mass it ever had at all previous snapshots. For level-1 subhalos, the PMF barely evolves over mass and redshift, and can be well fit with a double-Schechter function
\begin{equation}
        g_1(\mu)\equiv\frac{ \mathrm{d}N}{\mathrm{dln}\mu}=(a_1\mu^{\alpha_1}+a_2\mu^{\alpha_2})e^{-c\mu^d}.\label{eq:pmf}
\end{equation} Here, $\mu=m_{\rm peak}/M_{\rm host}$ is the ratio between the subhalo peak mass and the host halo mass. Following \cite{Jiang25}, we compute the host halo mass as the sum of the bound mass from all subhalos in a given FoF halo, to enhance the universality of the resulting PMF. The parameters to fit the level-1 PMFs across redshifts are $(a_1, a_2, \alpha_1, \alpha_2, c, d)=(0.0186, 0.2620, -0.9873, -0.5786, 5.491, 1.947)$. This result is consistent with the findings of \cite{Jiang25}, who found that the level-1 PMF is close to universal over halo mass, redshift and cosmology \rev{using the \hbt subhalo finder}. 

Starting from a universal level-1 PMF, \cite{Jiang25} also showed that the PMFs of level-$\ell$ subhalos, $g_\ell$, can be derived by self-convolutions of the level-1 PMF, 
\begin{equation}
        g_\ell(\mu)=\int^{\infty}_{\mu'=0} g_{\ell-1}(\mu')g_{1}(\beta\mu/\mu')\mathrm{d}\ln\mu'.
    \label{eq:PMF_conv}
\end{equation}
The additional parameter, $\beta$, accounts for the translation between the different mass definitions for subhalos and host halos. \cite{Jiang25} found a best-fitting $\beta=0.726$. Applying the model to Jiutian-300, we find that the best-fitting $\beta$ is slightly larger and evolves over redshift, with $\beta=0.877, 0.833, 0.813, 0.806$ at $z=0,1,2,3$, respectively. This evolving $\beta$ is needed particularly to match the PMFs at the low mass end ($\mu<10^{-3}$) which was not covered by the lower-resolution data in \cite{Jiang25}. As the level-1 PMF is consistent with being universal, the evolution in $\beta$ may be interpreted as reflecting an evolution in the total bound mass of the host halo for a given central mass at the peak mass time. 

Note even with an evolving $\beta$, the $z=0$ PMFs still show some deviations from the model at the low mass end. The level-1 PMF is slightly under estimated, while the level-2 and level-3 PMFs are over estimated. This may be understood as due to migration of the subhalo in levels. In \hbt, when a level-$\ell$ subhalo becomes unresolved, the level $\ell+1$ subhalos associated to it will be upgraded to level-$\ell$. This process could lead to an increase in the abundance of level-1 subhalos and decreases in that of higher-level ones. We leave detailed investigation and modelling of the deviation from universal PMFs to future works.


\subsection{galaxy light-cone catalogs}
\begin{figure*}
    \includegraphics[width=\textwidth, trim={0 4cm 2.5cm 2.5cm},clip]{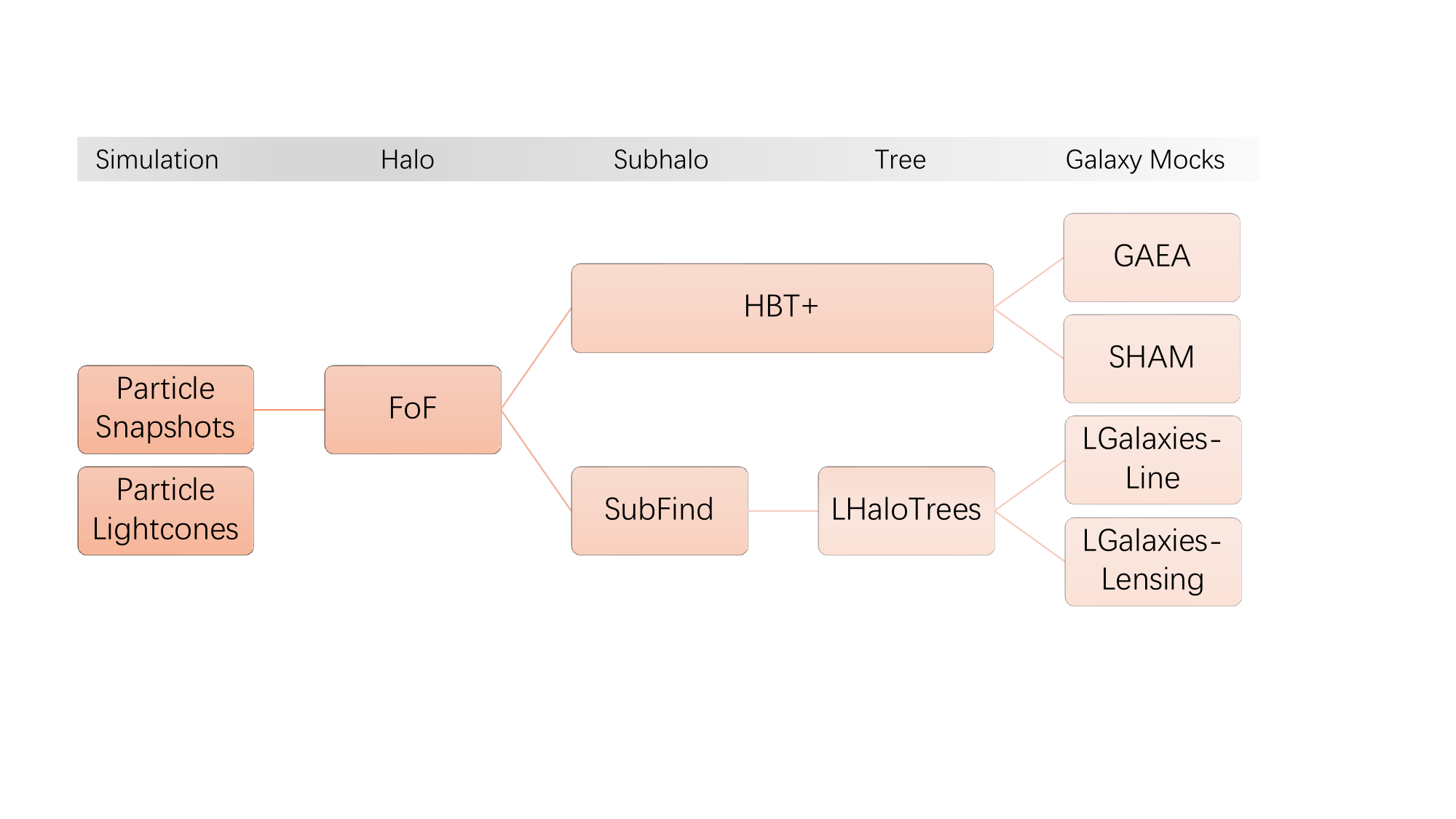}
    \caption{Flowchart of the post-processing pipelines and the corresponding data products for the primary runs. The simulation produces particle snapshots and particle lightcones. After running \fof halo finding on the snapshots, subhalos and merger trees are built with two pipelines, which are used to further produce four sets of mock galaxy lightcone catalogs.}\label{fig:flow}
\end{figure*}

To facilitate the application of the dark matter only simulations in galaxy surveys, it is essential to populate the simulations with galaxies using either physical or empirical models. In total, four sets of mock galaxy light-cone catalogs have been produced from the Jiutian primary runs, following different post-processing pipelines, as summarized in Figure~\ref{fig:flow}. \rev{Below, we first give a brief overview of the four mocks and their complementarity with each other, and provide more detailed descriptions about each mock in section~\ref{sec:gaea} to section~\ref{sec:sham}.}

Starting from the two sets of subhalo catalogs and merger trees, three independent versions of semi-analytical model (SAM) mock catalogs are produced, with different focuses. The \gaea-\hbt mocks are produced from the \gaea model~\cite{GAEAX20} applied to the \hbt catalogs, with additional \rev{improvements to the light-cone building}. The \lgalaxyp mocks \cite{Pei24} are produced by further developing the \lgalaxy model \cite{2015MNRAS.451.2663H}, for proper predictions on galaxy emission lines. Finally, a third SAM catalog is produced from the \lgalaxy model of \cite{Luo16}, with \rev{additional modelling of} the gravitational lensing effects on the galaxies (hereafter \lgalaxyl)~\cite{Wei+2018ApJ}.

Apart from the different focuses in the three SAM catalogs, the different models also have their own physical or technical advantages. For example, \gaea includes specific treatment on cold gas stripping and resolved mergers\footnote{When a subhalo sinks to the center of another subhalo and becomes phase mixed without experiencing tidal disruption, it is identified as a ``sink'' event by \hbt and represents a ``resolved'' merger between two subhalos.} between satellite galaxies according to the \hbt merger tree which are not considered in \lgalaxy. On the other hand, \lgalaxy has been developed to enable Bayesian inference of model parameters. Having these catalogs in parallel thus also provides the opportunity for assessing model uncertainties when using the mocks. 


\begin{figure*}
\begin{center}
    \includegraphics[width=0.75\textwidth]{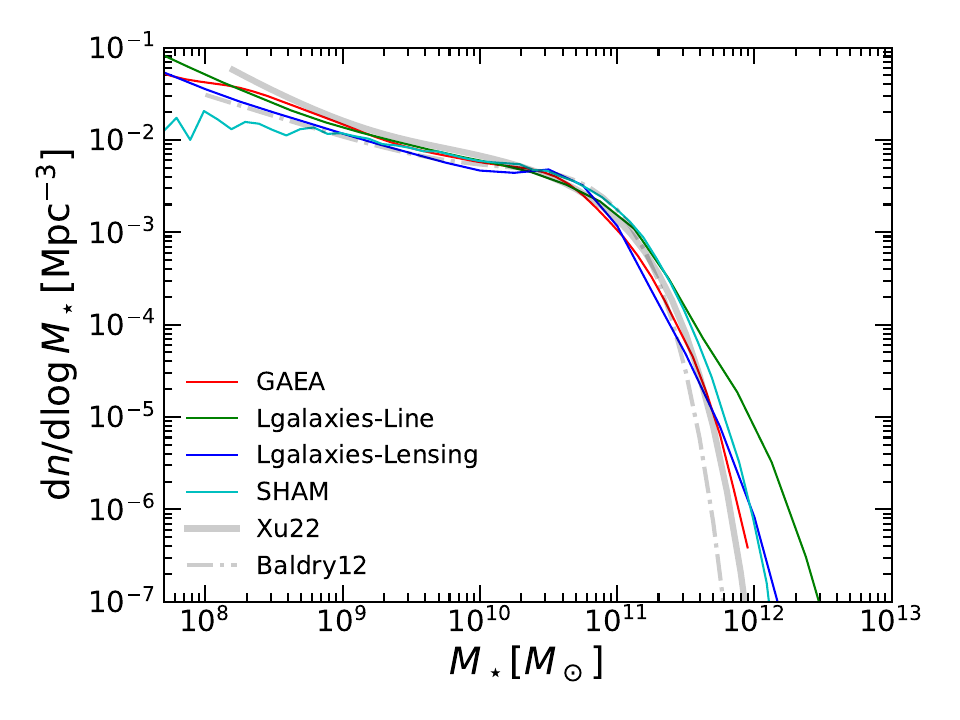}
    \end{center}
    \caption{Comparison of the $z=0$ stellar mass functions from the four mock catalogs of Jiutian-1G. Observational results from Xu et al. 2022 \cite{Xu2022} and Baldry et al. 2012 \cite{Baldry2012MNRAS} are also shown as labeled. 
    }
    \label{fig:smf}
\end{figure*}

In addition to the SAM mocks, an empirical mock based on SubHalo Abundance Matching (SHAM) is also produced\cite{Gu2024}, combining subhalos from Jiutian and galaxies from the DESI survey. SHAM sidesteps major uncertainties in the physical modeling of galaxy formation by exploiting the general correlation between the peak mass of a subhalo and \rev{the stellar mass or luminosity} of a galaxy. In this SHAM catalog, a real DESI galaxy is assigned to each selected subhalo in the simulation according to a high dimensional matching scheme, thus providing rich observational properties of the mock galaxies.

In Fig.~\ref{fig:smf}, the stellar mass functions from the four catalogs are compared with each other, and against observations. All of them can generally match the observed stellar mass functions from previous surveys at $z=0$. At the high mass end, the result from the \lgalaxyp is higher than observations. This is due to the weaker AGN feedback adopted in \lgalaxyp, in order to also match the stellar mass function at high redshifts, while the other two models are only tuned to match the $z=0$ stellar mass function. At the very low mass end, the galaxy catalog can be affected by limited resolutions of their host (sub)halos in the simulation, leading to the divergence in the predicted mass functions for $M_\star<4\times 10^8 M_\odot$. Despite the resolution effect, the result from the SHAM catalog significantly underestimates the observed stellar mass function at $M_\star<10^9\mathrm{M}_\odot$. This is because the SHAM mock is matched to the galaxy distribution from the DESI one-percent survey, which suffers from significant cosmic variance at this mass range~\cite{WangYR2024}.

\subsubsection{\gaea mocks}\label{sec:gaea}
\gaea \footnote{\url{https://sites.google.com/inaf.it/gaea}} is a semi-analytical galaxy formation model developed from \cite{delucia2007}. With decades of effort improving treatments on various physical processes \cite{delucia2014, hirschmann2016, xie2017, GAEAX20}, this rendition of \gaea \cite{GAEAX20} is particularly successful in reproducing the overall statistical properties of quenching processes and gas scaling relations. 

As is conventionally done in most SAM models, the original \gaea code \rev{first decompose the merger trees into independent ``forests" and then processes them forest by forest}. In this work, we adapt the \gaea code to work with \hbt outputs by evolving all galaxies in a given volume snapshot by snapshot. 
Galaxies in the output catalog share the same track ID as their host subhalos, so that the evolution history of each galaxy can be easily retrieved across time using the track ID. The output catalogs also provide links to galaxies in the same FoF group, and to the descendant galaxies that they will merge into. By tracing these links, the merger history of each galaxy can be easily built. 
Given that \hbt allows mergers between subhalos, the mergers between satellite-satellite galaxies are also considered in the galaxy model. The model parameters are tuned to match the $z=0$ stellar mass function measurement of \cite{Li09}.

The star formation rate of each galaxy, for both the disk and the bulge components, is saved at each snapshot from \gaea. These SFHs are then processed by \starduster \cite{starduster},\footnote{\url{https://github.com/yqiuu/starduster}} a stellar population synthesis code, to produce galaxy magnitudes. The unique advantage of \starduster is that it allows for fast computation of galaxy spectra and magnitudes with detailed modeling of dust obscuration. \rev{To achieve this, \starduster trains a neural network for dust modeling using simulations from the \textsc{SKIRT} code~\cite{SKIRT, SKIRT2, SKIRT3},\footnote{\url{https://skirt.ugent.be/root/_home.html}} a three-dimensional Monte Carlo radiative transfer code for simulating radiation in dusty astrophysical systems with adjustable geometry configurations.} 
Magnitudes are computed in a large number of filters adopted in existing and future surveys including the seven CSST photometric bands.

The outputs of \gaea at discrete snapshots are further processed with a light-cone builder, \blic\footnote{\url{https://github.com/zltan000/BLiC}}, to produce a continuous mock galaxy catalog as observed by a virtual observer. \blic works by first interpolating the trajectory of each galaxy recorded in the snapshots over time, and then solving for the time and position at which the galaxy intersects with the observer's past light-cone. \rev{\blic allows for different interpolation schemes to be used, including linear, cubic and spline interpolations in Cartesian coordinates, as well as linear interpolation in polar coordinates. When building the Jiutian lightcone, the trajectory of a galaxy is interpolated using a cubic polynomial, with constraints specified by the positions and velocities of the galaxy at the two bracketing snapshots.} The properties of the observed galaxy, including properties for its host subhalo, are also interpolated accordingly for outputs in the light-cone. 


Detailed descriptions of the Jiutian \gaea catalogs and the resulting galaxy light-cones, as well as forecasts for the CSST survey, will be presented in a separate paper (Tan et al., in preparation).

\begin{figure}[H]
    \centering
    \includegraphics[width=\linewidth]{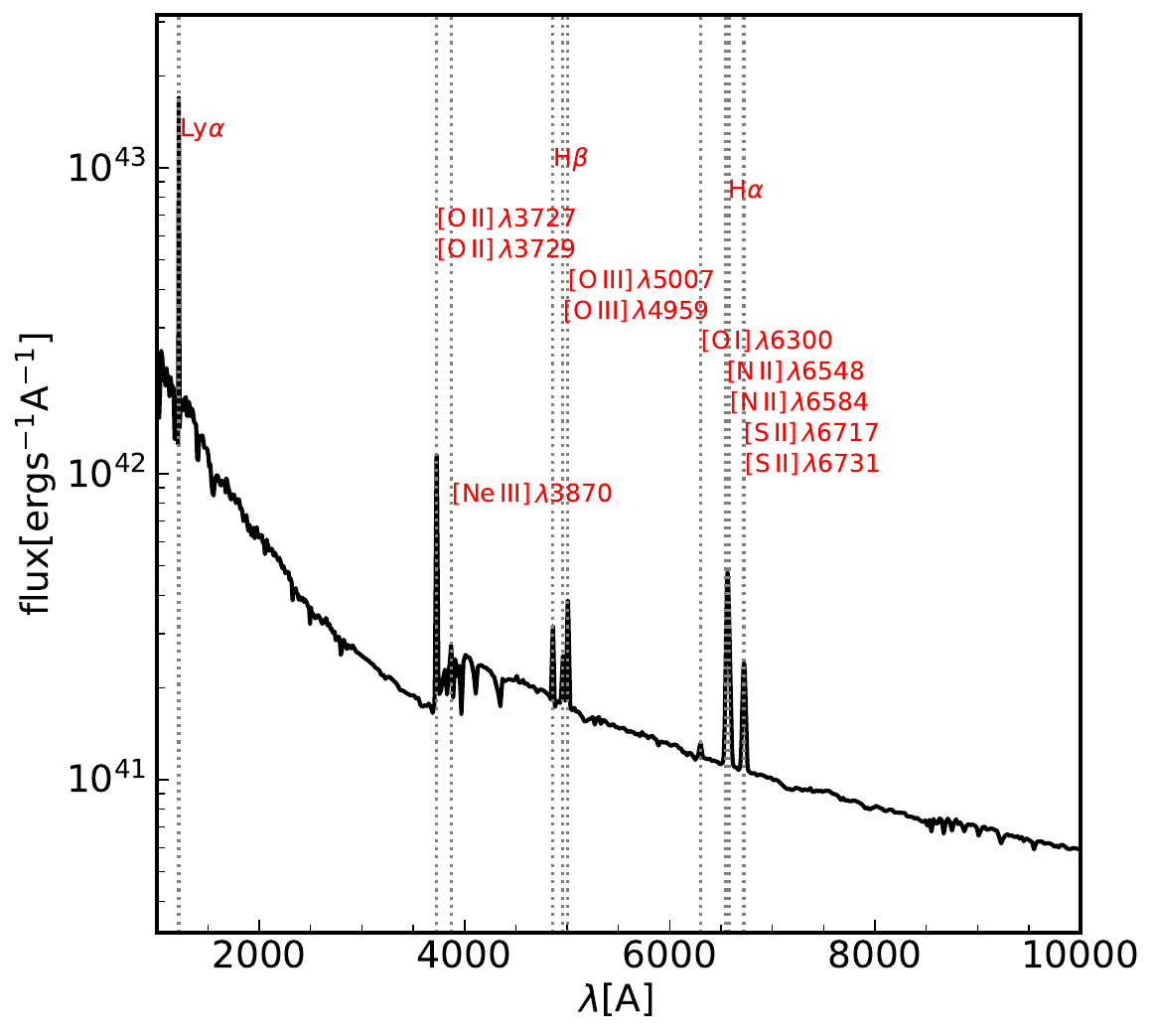}
    \caption{An example galaxy spectrum with emission lines predicted from the \lgalaxyp model. The host galaxy is a $M_\star=10^{10.88}\mathrm{M}_\odot$ late type galaxy. The individual emission lines are labeled in the plot.}
    \label{fig:emline}
\end{figure}

\subsubsection{Emission line mocks}
Emission lines carry important information on the properties of the interstellar medium (ISM) in a galaxy, while also providing crucial features for accurate determination of spectroscopic and photometric redshifts. Many recent and upcoming surveys are also targeting emission line galaxies as one of the main tracers of the large scale structure at $z\sim 1$. However, predictions on the emission line features are largely missing in previous SAMs (see e.g.,\cite{GALFORMemission, GAEAemissin} for recent efforts). 

In the Jiutian project, the \lgalaxy model has been extended to provide such predictions~\cite{Pei24}.
\lgalaxy was originally introduced by \cite{2001MNRAS.328..726S} and has undergone a series of updates in subsequent versions \cite{2004MNRAS.349.1101D,2006MNRAS.365...11C,2011MNRAS.413..101G,2015MNRAS.451.2663H}. In \cite{Pei24}, a new module is added to compute emission lines from the stellar \rev{continuum emission} and the properties of the ISM. An interpolation table of emission line ratios is first built from the photoionisation code \textsc{Cloudy}\cite{cloudy} for HII regions of different gas density, metallicity and ionisation state. This table is then coupled with the gas properties of each galaxy and the continuum emission computed from the stellar populations, to compute the luminosities of
13 most frequently used emission lines in the NUV and optical bands (see Fig.~\ref{fig:emline} for an example spectrum \rev{and table 3 of [41] for a list of the lines used}). This version also achieves better convergence over time resolution by improving the treatment of satellite disruption. Combined with \subfind merger trees of Jiutian-1G, \cite{Pei24} successfully reproduced various observational results, including the galaxy stellar mass function, emission line luminosity functions, and galaxy correlation functions within the redshift range of \(z = [0-3]\). A light-cone catalog covering 50 square degrees have also been produced from these mock galaxies.

\begin{figure*}
    \centering
    \includegraphics[width=\textwidth]{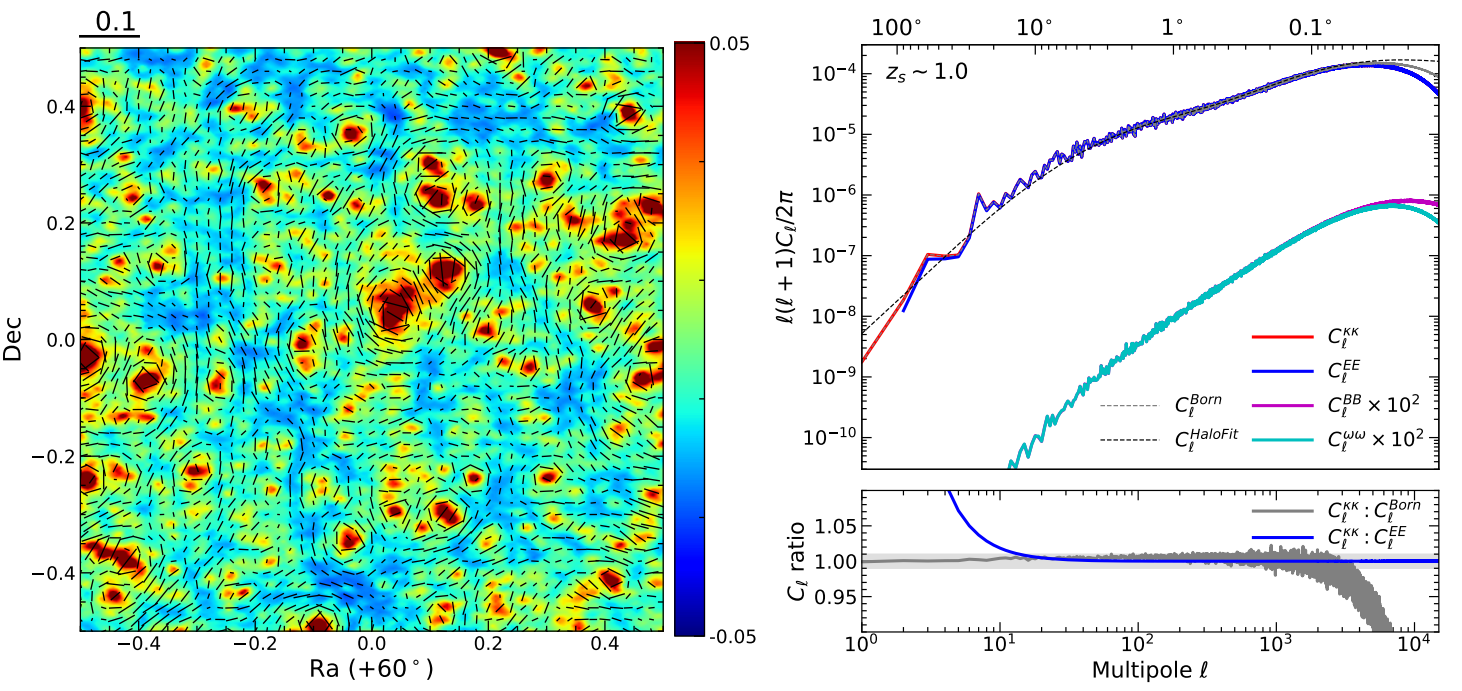}
    \caption{\textit{Left}: The convergence map (colour filled contour) from weak lensing simulation for sources at $z_{\rm s} = 1.0$ within a $1 \times 1 {\rm deg}^2$ region and the overlaid sticks indicate the lensing shear signal. \textit{Right}: The angular power spectrum of the convergence (red solid line), shear E-mode (blue), B-mode (magenta), and rotation mode (cyan) for sources at $z_s = 1.0$. The predictions from the Born approximation (gray dashed line) and revised Halofit model \cite{Takahashi+2012ApJ_Halofit} (black dashed line) are also presented for comparison. The lower panel shows the ratio of the angular power spectrum between the lensing convergence and the shear E-mode, with the gray shaded region indicating a 1\% deviation. }
    \label{fig:lens0}
\end{figure*}

\subsubsection{Lensing mocks}
The lensing mocks are designed to provide high-precision simulation of the gravitational lensing effects on galaxies. These mocks are essential for testing and calibrating the lensing analysis pipelines, as well as for studying the impact of lensing on galaxy clustering and cosmology.

To generate the lensing mocks, we first apply the \lgalaxyl semi-analytical model to the Jiutian-1G simulation to produce mock galaxies at each snapshot. This model has been extended to include the shape and alignment of each galaxy, assigned according to a statistical model derived from Wei et al. 2018 \cite{Wei+2018ApJ}. Galaxy and particle lightcones are then constructed up to redshift $z_{\rm max} \sim 3.5$ for a virtual observer at $z=0$, ensuring consistency between the underlying dark matter and the galaxy distributions. Weak lensing effects are simulated using a multi-plane ray-tracing algorithm on the curved sky \cite{Wei+2018ApJ, sFMM}, which propagates light through the simulated cosmic structure to generate convergence and shear maps with arcminute resolution. In the left panel of Fig.~\ref{fig:lens0}, we show a slice of the simulated convergence map for sources at redshift $z_{\rm s} = 1.0$ within a $1 \times 1 {\rm deg}^2$ region. As expected, tangential shears (sticks) can be observed around the convergence peaks. Correspondingly, the angular power spectrum statistics for convergence (red solid line), shear E-mode (blue) and B-mode (magenta), and rotation mode (cyan) \rev{(see e.g., \cite{Becker13} for definitions)} with the sources at $z_s = 1$ are presented in the right panel. On weak lensing scales, the measured power from the ray-tracing simulation shows good agreement with the prediction from the Born approximation, with a relative error of less than 10\% at $\ell \le 6000$. As predicted by theory, at small scales (high-$\ell$), shear E-mode power matches the convergence power, while the deviation  at large scales (low-$\ell$)  arises from an additional factor of $(\ell-1)(\ell+2)/\ell/(\ell+1)$. 

\begin{figure}[H]
    \centering
    \includegraphics[width=0.5\textwidth]{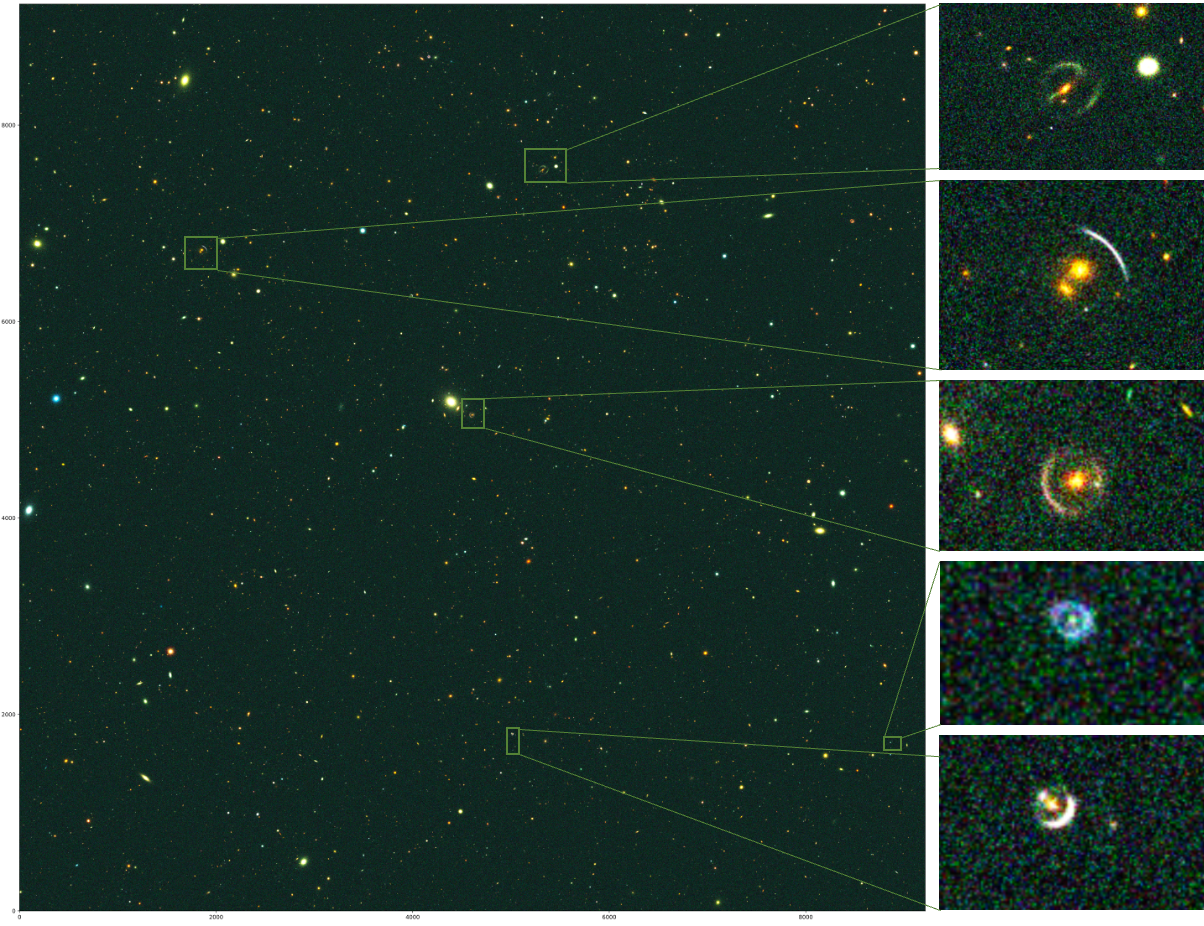}
    \caption{An illustration of the lensing mock catalog in a $11' \times 11'$ field, with several simulated strong lensing events shown in zoom-in panels. \rev{From top to bottom, the lens galaxies have stellar masses and redshifts of $(\log(M_\star/[\msunh]), z)=(11.6, 0.76)$, $(12.1, 1.05)$, $(11.4, 0.67)$, $(10.5, 1.05)$, $(10.9, 0.85)$}.}
    \label{fig:lensmock}
\end{figure}

To model strong lensing, it is necessary to directly simulate image distortions instead of only computing the shears. To this end, we select a $1\deg^2$ field from the weak lensing lightcone, and adapt the Pipeline for Images of Cosmological Strong-lensing (PICS) \cite{PICS_Li+2016ApJ} to generate synthetic images for CSST observations and simulate strong lensing distortions. Based on the resolution of the CSST survey, galaxies with Einstein radius $\theta_{\rm E} > 0.2''$ were selected to generate strong lensing images, \rev{corresponding to $\sim 1.4$ strong lenses per $\mathrm{arcmin}^2$}. Galaxy images are first modeled by Sersic profiles with the galaxy properties recorded in the lightcone catalog, and lensed images are produced \rev{using PICS assuming a singular isothermal ellipsoid lens profile.}
Foreground stars, instrumental noise and artifacts, and the point spread function of CSST are all accounted for by the pipeline.  In Fig.~\ref{fig:lensmock} we show a $11'\times 11'$ cutout of the simulated field, with zoomed images of several strong lensing systems.

The resulting weak lensing catalogs contain key observables such as galaxy shapes, magnification factors, and shear components in addition to classical observables from the \lgalaxy model, allowing for direct comparisons with CSST survey data. Together with the strong lensing simulations, the lensing mocks have been used to support the development of CSST imaging simulator (Wei et al. in prep.) and to forecast the scientific potential of the CSST weak lensing survey \cite{Zhang+2024A&A, Li+2024AJ, Qu+2023MNRAS}.

\subsubsection{SHAM mocks}\label{sec:sham}
The SHAM mocks~\cite{Gu2024} are produced by populating the dark matter subhaloes in the Jiutian-1G simulation with galaxies in the DESI survey. 
First, \hbt subhaloes from different snapshots, including disrupted ones traced by the most bound particle, are combined into a lightcone. Each subhalo is then assigned a luminosity according to the $z$-band luminosity function measured from the DESI one-percent spectroscopic survey of the DESI early data release~\cite{DESI_EDR}. The assignment is done to ensure that the brightest galaxies are on average assigned to the most massive subhaloes (in peak mass) at the same abundance level, with an additional scatter \rev{of $\sigma_{\mathrm{log}(L_z)}=0.15\,\mathrm{dex}$ at a fixed peak mass according to the conditional luminosity measurement of \cite{Yang08}}. Starting from the redshift, assigned luminosity and halo mass of each subhalo, a real galaxy from the DESI Legacy Survey (LS) Data Release 9~\cite{DESI_LS} that is closest in the three properties is matched to the subhalo. To make this matching, the observed halo mass of each DESI galaxy is estimated from an updated version of the halo-based group catalog of \cite{Yang2021}. The resulting observationally matched catalog thus contains all the galaxy properties that are available from the DESI LS. \rev{A small fraction of $0.8\%$ of galaxies also have matched spectra from the one percent survey.}
The image of each matched galaxy is also extracted from DESI LS and synthesized on the sky to form a complete mock imaging survey. Survey geometry and magnitudes cuts from the CSST and DESI LS are also applied to produce mocks for the corresponding surveys. \rev{As limited by the depth of the DESI LS catalog used in the matching, the current mock catalog is complete down to a z-band magnitude of $m_z=21$ and covers the redshift range of $0<z<1$.}

These mock catalogs can well match the galaxy distributions in the DESI LS and spectroscopic surveys. They allow for end-to-end comparisons between the mocks and observations, and can be particularly useful for testing various selection effects and analysis pipelines.

\section{Emulator runs}\label{sec:emulator}
The emulator runs, also known as the \textsc{Kun} simulations~\cite{emulator}, are designed to explore the parameter space of the concordance cosmology and to provide a fast and accurate way to predict cosmological observables for a wide range of cosmological parameters. 

\begin{table*}[t]
    \centering
    \begin{threeparttable}
    \caption{Parameter space of the emulator runs}
    \begin{tabular}{ccccccccc}
    \toprule
    parameter\tnote{1)} & $\Omega_b$ & $\Omega_{cb}$ & $H_0 [\mathrm{km\,s^{-1}\,Mpc^{-1}}]$ & $n_s$ & $A_s\times 10^9$ & $w_0$ & $w_a$ & $\sum m_\nu [\mathrm{eV}]$ \\	 	 	 	 	 	 	
    Min & 0.04 & 0.24	& 60	& 0.92	&1.7	&-1.3&	-0.5&	0\\
Max	& 0.06 &	0.40	& 80	&1.00	&2.5	&-0.7&	0.5&	0.3\\
\bottomrule
    \end{tabular}
    \label{tab:emulator_run}
    \begin{tablenotes}
    \footnotesize
       \item[1)] The listed cosmological parameters are the $z=0$ baryon and matter (without massive neutrinos) density parameters and the Hubble constant, ($\Omega_b, \Omega_{cb}, H_0$), the primordial power index and amplitude, $(n_s, A_s)$, the parameters for an evolving dark energy equation of state, $w_{\rm DE}(a)=w_0+w_a(1-a)$, and the sum of neutrino masses, $\sum m_\nu$.  
    \end{tablenotes}
    \end{threeparttable}
\end{table*}

To this end, we vary eight cosmological parameters 
around the Planck 2018 cosmological model, allowing for a dynamical dark energy equation of state and non-zero neutrino masses (see Table~\ref{tab:emulator_run}). The Sobol sequence sampling \cite{sobol1967distribution} method is adopted to sample the 8-dimensional parameter space efficiently at $129$ points including the fiducial cosmology of Planck 2018. For each of the sampled cosmologies, we run a $N$-body simulation of $3072^3$ particles in a cubic box of $1\,\gpch$, corresponding to a particle mass of $2.87 \frac{\Omega_{cb}}{0.3} \times 10^{9}\, h^{-1}M_{\odot}$.
Here, $\Omega_{cb} = \Omega_{m} - \Omega_{\nu}$ represents the matter density without the massive neutrino component at present.

The simulations are run using a modified version of the \gadget-4 code\footnote{\url{https://github.com/czymh/C-Gadget4}} for the generalized cosmological model including dynamical dark energy and massive neutrinos. We evolve the cold matter particles only and incorporate the impact of massive neutrinos through a gauge transformation \cite{Nmgauge} of the particle coordinates. To reduce storage requirements, we only output 12 snapshots between $z=3$ and $0$. Halos and subhaloes are identified using the built-in \fof and \subfind methods. An alternative halo and subhalo catalog is produced using the \textsc{Rockstar} \cite{rockstar} phase-space halo finder. For each of the simulations, we also output particle light-cones between $z=3$ and $0$ \rev{on the fly}, over two regions of \rev{46$\times$46} square degrees each. The line of sight of the pyramid-like cone is selected exquisitely to reduce the box replication effect \cite{2024MNRAS.534.1205C}. \rev{Full sky} density and convergence maps 
are also produced \rev{on the fly} at every $50\,\mpch$ comoving distance interval to facilitate gravitational lensing studies. 

For each of the \textsc{Kun} simulations, we compute a number of cosmological statistics including various power spectra, halo mass function, halo-halo and halo-matter correlation functions, and weak lensing convergence power spectrum. These outputs are then used to train an emulator that uses Gaussian Process Regression to interpolate between them and predict the statistics for any set of cosmological parameters within the training range. \rev{More detailed descriptions of the simulations and the resulting emulator are provided in \cite{emulator}. Below we briefly summarize the performance of the emulator.}

\begin{figure}[H]
    \centering
    \includegraphics[width=0.45\textwidth]{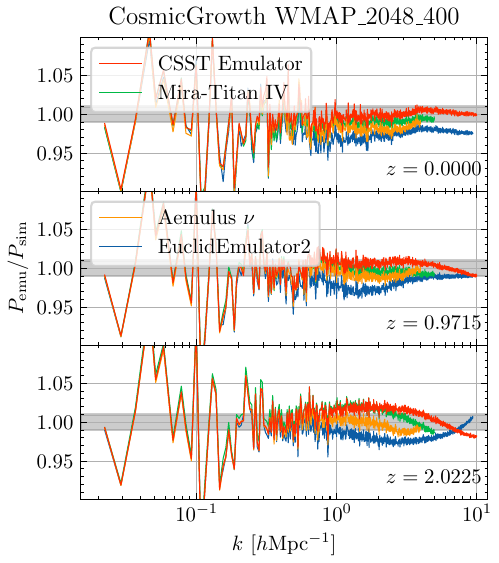}
    \caption{Predictions of the matter power spectra from the CSST Emulator, Mira-Titan IV, Aemulue-$\nu$, and EuclidEmulator2 emulators, compared against the WMAP\_2048\_400 simulation in the \textsc{CosmicGrowth} suite. Figure reproduced from \cite{emulator}. }
    \label{fig:emulator}
\end{figure}

\rev{The emulator performance is assessed using both the Kun simulations and a number of external simulations, including Quijote\cite{Quijote}, FLAMINGO\cite{FLAMINGO}, AbaccusSummit\cite{AbacusRun} and CosmicGrowth\cite{CosmicGrowth}.}
The resulting CSST Emulator\footnote{\url{https://github.com/czymh/csstemu}} can achieve 1\% precision in predicting the real-space power spectrum for \( k \leq 10 \, h{\rm Mpc}^{-1} \) and redshift \( 0 \leq z \leq 2 \) across the entire cosmological parameter space. The performance at higher redshifts \( 2 < z \leq 3 \) remains at \( k\leq 5\, h{\rm Mpc}^{-1} \) while becoming slightly worse for smaller scales due to \rev{shot noise that starts to dominate the signal at those scales}. The same precision is achieved for \( k \leq 5 \, h{\rm Mpc}^{-1} \) across all redshifts in redshift space. The real-space correlation function can be predicted to 1\% precision for \( 10^{-2} \, h{\rm Mpc}^{-1} \leq r \leq 50 \, h{\rm Mpc}^{-1} \) and redshift \( 0 \leq z \leq 3 \). The weak lensing convergence field power spectrum can achieve 1\% precision predictions for \( \ell \leq 10^4 \) and source galaxy redshift \( 0.5 \leq z_s \leq 3 \).
Tests show that the emulator's precision surpasses that of the Mira-Titan emulator~\cite{MiraTitanIV} and the EuclidEmulator2~\cite{EuclidEmulator2}. 
Figure~\ref{fig:emulator} shows an example of comparing the performance of the CSST Emulator, Mira-Titan IV, Aemulue-$\nu$\cite{AemulusNv}, and EuclidEmulator2 emulators, using the WMAP cosmology of the high-resolution \textsc{CosmicGrowth} simulation\cite{CosmicGrowth}. 
For all redshifts, the CSST Emulator is able to accurately predict the simulation power spectra up to $k=5\, h{\rm Mpc}^{-1}$, while the EuclidEmulator2 predictions start to perform worse at $k=1\,h{\rm Mpc}^{-1}$ and deviate by $\sim 3\%$ at $k=2\,h{\rm Mpc}^{-1}$. \rev{The deviation of the EuclidEmulator2 is probably related to the limited resolution of the simulations used to train this emulator. Even though a correction factor has been introduced in their emulator paper\cite{EuclidEmulator2}, it is not yet implemented in the current version of EuclidEmulator2.} 
The Mira-Titan IV performs slightly worse than the CSST Emulator at $z=0$.
The most recent Aemulus-$\nu$ emulator has a similar accuracy with the CSST Emulator, while our emulator extends the usable range from \( 5\, h{\rm Mpc}^{-1}\) to \(10\, h{\rm Mpc}^{-1}\). 

\section{Reconstruction runs}\label{sec:resim}
The reconstruction runs focus on reproducing the evolution of the Universe using initial conditions reconstructed from existing observations. Unlike conventional simulations that can only match the observed Universe in the statistical properties, reconstruction simulations are designed to match the density and optionally the velocity field of the Universe, by essentially trying out different initial conditions in an efficient way. Eventually we aim to reconstruct the initial condition of the Universe using CSST observations. As such observations are not yet available, at this stage we work with initial conditions constructed from the SDSS observations. In particular, the ELUCID project~\cite{ELUCID, ELUCID3} has reconstructed the initial conditions from SDSS using a group-based estimation of the cosmic density field and a Hamiltonian Mento-Carlo regression method. In the Jiutian project, we advance the ELUCID simulation to better resolutions by performing a set of zoom-in hydrodynamical simulations of certain regions in SDSS, including the Coma cluster and a large low-density region~\cite{Resim1, Resim2}.

To perform these simulations, the region of interest is first identified in the ELUCID simulation at $z=0$, and resampled with a higher resolution in the initial condition, followed by consecutively lower resolution regions outside. Gas particles are also generated in the high density region. The simulation is then re-run with the enhanced initial condition using the \textsc{Simba} code~\cite{Simba}. A number of different feedback models are adopted to explore the influence of different baryonic physics on the final structure, including i) the fiducial \textsc{Simba} model as in the original code; ii) a modification of the fiducial model with enhanced star formation and supernova feedback, and weakened AGN feedback; iii) enhanced star formation and supernova feedback, enhanced AGN feedback; iv) no AGN feedback. 

For the Coma cluster region, the four models are evolved from the same initial condition, following the same cosmology (WMAP5, \cite{WMAP5}) as the original ELUCID simulation. The high resolution region is sampled with particles of masses $3.2\times 10^7\msunh$ and $6.58\times 10^6\msunh$ for dark matter and gas, respectively, with a gravitational softening length of $0.8\kpch$. In addition, we have also run two simulations (adopting model i and iv) for a void region in the SDSS at $z\approx 0.05$, representing a different environment from the cluster region. This void region is resolved with a relatively mass lower resolution of $5.4\times 10^7\msunh$ in gas particle mass and a softening of $1.6\kpch$. For fair comparisons with the void simulation, models i to iii are also simulated at the same resolution for the Coma cluster. In total, 9 simulations are produced. Detailed information of these simulations is presented in \cite{Resim2}. 

\begin{figure*}[htp]
    \includegraphics[width=\textwidth]{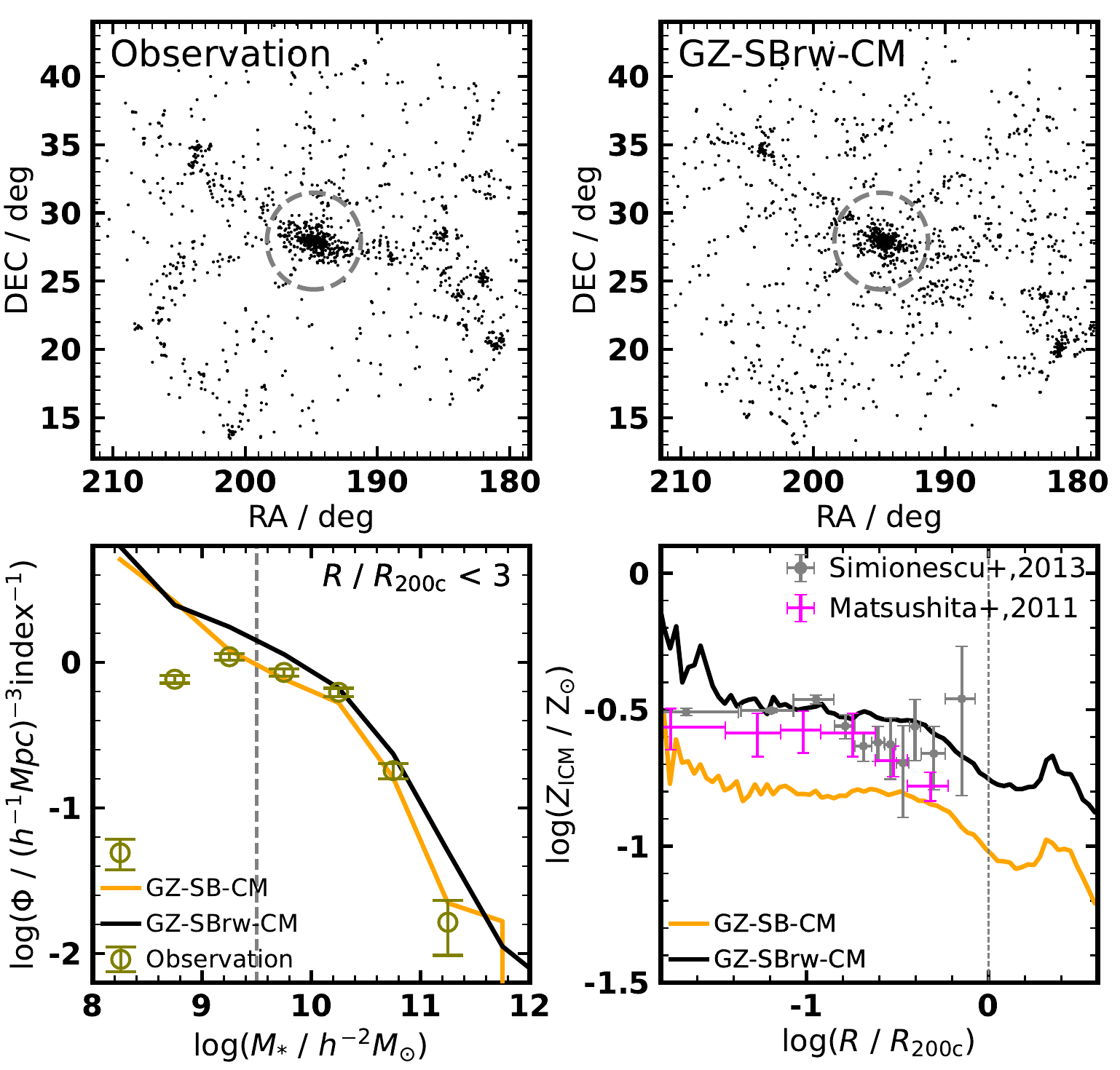}
    \caption{Comparisons between observations and model predictions around the Coma galaxy cluster are presented. The top two panels show the two-dimensional distributions of galaxies in the J2000.0 coordinate system(left: SDSS, right: model). Only galaxies with $\log M*/h^{-2}M_{\odot}>9,5$ within a radius of 20 $Mpc/h$ are included. The bottom left panel illustrates the stellar mass function for galaxies located within three times the virial radius of the Coma cluster, as denoted by the circle in the upper panels . The vertical gray dashed line represents the estimated complete mass limit of the observation. The bottom right panel depicts the ICM metallicity as a function of radius to the Coma cluster. GZ-SB-CM represents the fiducial SIMBA model, while GZ-SBrw-CM represents the SIMBA model with enhanced star formation/supernova feedback and weakened AGN feedback. It is evident that the GZ-SBrw-CM model aligns with the observations more closely than the GZ-SB-CM model.}\label{fig:coma}
\end{figure*}

These simulations provide detailed predictions on the evolution histories of various matter components in the Coma cluster and the void region. By construction, they reproduce the final largescale structure around the simulated regions in the real Universe (see Fig.~\ref{fig:coma} for an example). This makes the comparisons between the simulation and observations free from cosmic variance. As a result, these reconstruction runs are particularly useful for studying the properties of galaxies in different environments, and for assessing the influences of baryonic processes on galaxy formation in a "one-to-one" manner.

\section{Extension runs}\label{sec:extension}
The nature of dark matter, dark energy and the masses of neutrinos are among the major science goals of next stage cosmological surveys including CSST. To help addressing these key questions, we carry out four sets of extension runs, covering models of warm dark matter, modified gravity interpretation of dark energy, interacting dark energy and different neutrino masses.


Table~\ref{tab:extension} lists the simulations included in this module. Each category also includes a \LCDM run using the same code and numerical specifications to facilitate comparative studies.

\begin{table*}[t]
\footnotesize
\begin{threeparttable}
\caption{Jiutian extension runs.}\label{tab:extension}
\doublerulesep 0.1pt \tabcolsep 13pt 
\centering
\begin{tabular}{l l c c c c c}
\toprule
\hline
 & Model & Boxsize & Particles & Particle Mass ($\msunh$) & Snapshots & Software \\
 \midrule 
 \multirow{4}{*}{WDM} & \LCDM \tnote{1)} & \multirow{4}{*}{100Mpc/h} & \multirow{4}{*}{2048$^{3}$} & \multirow{4}{*}{$1.02\times10^{7}$} & 163 & \multirow{4}{*}{\gadget4} \\
 & 3 keV &  &  &  & 178 &  \\
 & 1.2 keV &  &  &  & 117 &  \\
 & 0.5 keV & & & & 178 & \\
\midrule
\multirow{6}{*}{Neutrino} & \LCDM \tnote{2)} & \multirow{3}{*}{1.5Gpc/h} & \multirow{3}{*}{2048$^{3}$} & \multirow{3}{*}{$3.40\times10^{10}$} & \multirow{3}{*}{38} & \multirow{6}{*}{\cube} \\
 & 0.05eV &  &  &  &  &  \\
 & 0.1eV &  &  &  &  &  \\
\cline{2-6}
 & \LCDM \tnote{2)} & \multirow{3}{*}{3Gpc/h} & \multirow{3}{*}{2048$^{3}$} & \multirow{3}{*}{$2.72\times10^{11}$} &\multirow{3}{*}{38} &  \\
 & 0.05eV &  &  &  &  &  \\
 & 0.1eV &  &  &  &  &  \\
\midrule
\multirow{4}{*}{$f(R)$} 
 & \LCDM \tnote{2)} & \multirow{4}{*}{3Gpc/h} & \multirow{4}{*}{2048$^{3}$} & \multirow{4}{*}{$2.74 \times 10^{11}$} & \multirow{4}{*}{40} & \multirow{4}{*}{\textsc{Ecosmog}} \\
  & F4 &  &  &  &  &  \\
 & F5 &  &  &  &  &  \\
 & F6 &  &  &  &  &  \\
\midrule
\multirow{7}{*}{IDE} & \LCDM \tnote{2)} & \multirow{7}{*}{1.5Gpc/h} & \multirow{7}{*}{2048$^{3}$} & $3.42 \times 10^{10}$ & 47 & \multirow{7}{*}{\textsc{ME-Gadget4}} \\
 & Conformal1 &  &  & variable\tnote{3)} & 47  &  \\
 & Conformal2 &  &  & variable & 47 &  \\
 & Disformal1 &  &  & variable & 39 &  \\
 & Disformal2 &  &  & variable & 39 &  \\
 & Quintessence1 &  &  & $3.42 \times 10^{10}$ & 48 &  \\
  & Quintessence2 &  &  & $3.42 \times 10^{10}$ & 48 &  \\
\hline
\bottomrule
\end{tabular}
\begin{tablenotes}
\item[1)] Planck 2015 cosmology~\cite{Planck15}
\item[2)] Planck 2018 cosmology~\cite{Planck2018}, same as in the Jiutian primary runs
\item[3)] The particle mass is variable due to dark matter and dark energy interaction. 
\end{tablenotes}
\end{threeparttable}
\end{table*}
\subsection{Warm dark matter}
Warm dark matter is a \rev{popular alternative } to cold dark matter, as they produce almost identical large scale structure in the Universe and only differ on small scales. 
The warm dark matter simulations, also known as Kanli simulations in \cite{Kanli}, consist of three WDM runs with different WDM particle masses, in addition to one CDM run. They are conducted with the \gadget-4 code, adopting the cosmological parameters of \cite{Planck15}, with $\Omega_m=0.3156$, $\Omega_{\Lambda}=0.6844$, $h=0.6727$, $n_s=0.967$ and $\sigma_8=0.81$. Each simulation resolves $2048^3$ particles in a cubic box of $100 h^{-1}\mathrm{Mpc}$, corresponding to an $N$-body particle mass of $1.02\times10^{7} h^{-1} M_{\odot}$, with a gravitational softening of $\epsilon=1 h^{-1}\mathrm{kpc}$. 

We adopt the capacity constrained Voronoi tessellation (CCVT) method to create the pre-initial particle loads \cite{2018MNRAS.481.3750L} for initial conditions, and ignore initial thermal velocities which have negligible effects on the structures at present day. The initial power spectra of the WDM models are related to the CDM model through 
\begin{equation}
    P_{\mathrm{WDM}}(k)=T^2(k) P_{\mathrm{CDM}}(k),
\end{equation}
with a transfer function \cite{2001ApJ...556...93B},
\begin{equation}
    T(k)=\left[1+(\theta k)^{2 \nu}\right]^{-5 / 1.2}.
\end{equation}
The parameter $\theta$ is determined by the thermal relic WDM particle mass, $m_\chi$, as
\begin{equation}
    \theta=0.05\left[\frac{m_{\chi}}{1 \mathrm{keV}}\right]^{-1.15}\left[\frac{\Omega_{\chi}}{0.4}\right]^{0.15}\left[\frac{h}{0.65}\right]^{1.3} h^{-1} \mathrm{Mpc}.
    \label{eq:theta}
\end{equation}
Here $\Omega_{\chi}$ is the WDM density parameter. Three thermal relic WDM particle masses, $m_{\chi}=3.0, \ 1.2,\ 0.5\rm{keV}$, are adopted in our simulations. Note these WDM masses have been largely ruled out by current observations~\cite{Enzi21, Nalder21}. Despite this, they are adopted to enable theoretical studies of the $m_\chi$ dependence of the cosmic structure by having detectable differences at the adopted resolution. 

\begin{figure*}
\includegraphics[width=\textwidth]{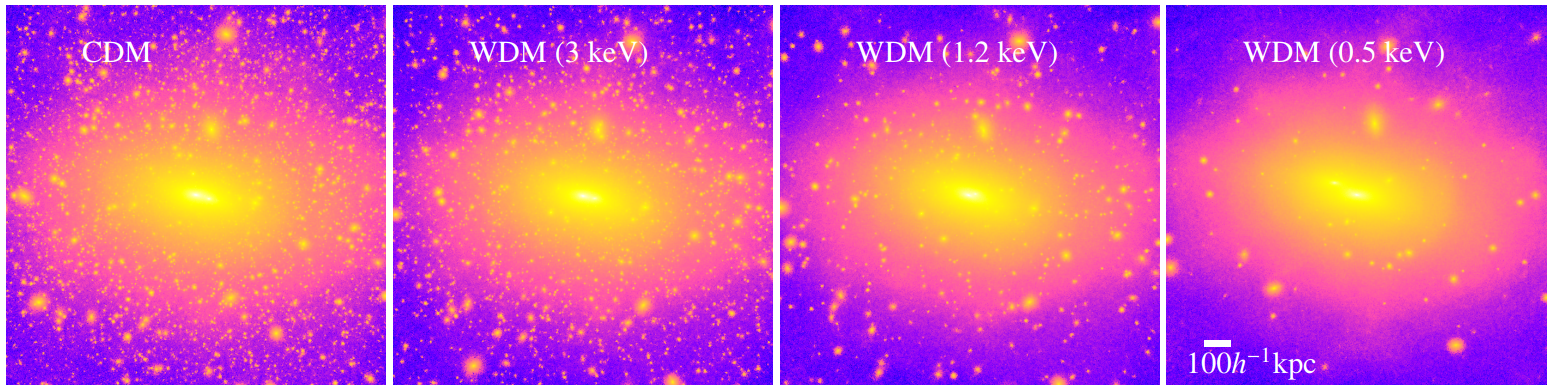}
    \caption{Projected images of the same cluster halo under different dark matter particle models from the WDM runs. The images are created by projecting the squared density of dark matter particles in the \fof halo in a box of $1.44\,\mpch$ per side.}\label{fig:WDM}
\end{figure*}

For each of the simulations, \fof halos and \hbt subhalos are identified. For the WDM runs, the subhalos are further processed following \cite{Lovell14} to identify spurious objects due to numerical noise. In Figure~\ref{fig:WDM}, we show the projected images of an example cluster in the four runs, which reveals the different amount of substructures across the simulations. These simulations have been used to derive an analytical model for the spatial and mass distribution of subhalos\cite{Kanli}, $P(m, m_{\rm acc}, r|M, m_{\chi})$, with $m$, $m_{\rm acc}$, $r$ being the final mass, peak mass and halo-centric distance of a subhalo in a host halo of mass $M$, generalizing the unified subhalo model of \cite{Han16} to WDM cosmologies with arbitrary $m_{\chi}$.

\subsection{$f(R)$ gravity}
Another extension is the exploration of modified gravity models. One of the most popular models is the so-called $f(R)$ gravity model. This model modifies Einstein's general relativity by replacing the Ricci scalar $R$ with an arbitrary function of $R$. The functional form of $f(R)$ is chosen so that it gives rise to the late-time cosmic acceleration on large scales without violating the solar system constraints. One of the viable models is the Hu-Sawicki model \cite{Hu-Sawicki}:
\begin{equation}
f(R) = -m^2\frac{c_1\left(-R/m^2\right)^n}{c_2\left(-R/m^2\right)^n+1},
\label{eq:hu_sawicki}
\end{equation}
where $m^2\equiv8\pi G\bar{\rho}_{\rm m,0}/3=H_0^2\Omega_{\rm m}$,  $\bar{\rho}_{\rm m,0}$ is the mean background matter density, $H_0$ is the Hubble expansion rate today and $\Omega_{\rm m}$ is the matter density parameter. In the high curvature regime $-\bar{R}\gg m^2$ with \rev{$\bar{R}$ being the background curvature}, we have $f(\bar{R})\approx-m^2c_1/c_2$ which is a constant. 

\rev{To yield a background expansion history close to that of the $\Lambda$CDM model}, we can choose $c_1/c_2=6\Omega_{\Lambda}/\Omega_{\rm m}$, where $\Omega_\Lambda\equiv1-\Omega_{\rm m}$, such that (see \cite{Hu-Sawicki} for detailed derivations),
\begin{equation}
-\bar{R} = 3m^2\left(a^{-3}+4\frac{\Omega_\Lambda}{\Omega_{\rm m}}\right) \approx 3m^2\left(a^{-3}+\frac{2}{3}\frac{c_1}{c_2}\right),
\label{eq:R}
\end{equation}
\rev{where $a$ is the expansion factor of the background universe.}
\rev{The derivative of the $f(\bar{R})$ function for the background universe, $\bar{f}_R\equiv{\rm d}f(\bar{R})/{\rm d}\bar{R}$, can be found as}
%
\begin{equation}
\bar{f_R} = -\frac{c_1}{c_2^2}\frac{n\left(\frac{-\bar{R}}{m^2}\right)^{n-1}}{\left[\left(\frac{-\bar{R}}{m^2}\right)^n+1\right]^2} \approx -n\frac{c_1}{c_2^2}\left(\frac{m^2}{-\bar{R}}\right)^{n+1},
\label{eq:scalar_field}
\end{equation}
in which,
\begin{equation}
\frac{c_1}{c_2^2} = -\frac{1}{n}\left[3\left(1+4\frac{\Omega_{\Lambda}}{\Omega_{\rm m}}\right)\right]^{n+1}\bar{f}_{R0}.
\label{eq:c1/c22}
\end{equation}
Note that $\bar{f}_{R0}$ denotes the present-day background value of $\bar{f}_R$. 

\rev{In the quasi-static and weak-field limits, the modified Poisson equation for Newtonian potential $\Phi$ and the field equation for $f_R\equiv{\rm d}f(R)/{\rm d}R$ are given by  }
\begin{align}\label{eq:poisson}
\nabla^2 \Phi &= 4 \pi G  a^2 \delta \rho_m - \frac{1}{2} \nabla^2 f_R, \\
\nabla^2 f_R &= -\frac{a^2}{3} \delta R - \frac{8 \pi G a^2}{3} \delta \rho_m,
\end{align}
\rev{where $\delta \rho_m=\rho_m-\bar{\rho}_m$ and $\delta R=R-\bar{R}$ are the matter density perturbations and the perturbation of the Ricci curvature respectively. $\delta R$ can be written as a function of $f_R$ and $\bar{f}_R$, which plays the role of the potential for $f_R$. The above two equations are numerically solved in our simulations.}

We adopt the values $n=1$ and $|\bar{f}_{R0}|=10^{-4}, 10^{-5} \text{ or } 10^{-6}$ (F4, F5 or F6, respectively) in this work. We run simulations for F4, F5 or F6, respectively, to study their impact on the formation of cosmic structures and the distribution of galaxies. These simulations help us understand how modified gravity affects the observables that will be measured by the CSST. The simulations are conducted using \textsc{Ecosmog} \cite{ECOSMOG}, a modified gravity extension of the N-body simulation code \textsc{Ramses} \cite{RAMSES} that efficiently models the effects of $f(R)$ gravity. 


\subsection{Interacting Dark Energy (IDE)}
We also explore the interacting dark energy (IDE) model, where dark energy interacts with dark matter, leading to a transfer of energy between the two components. This interaction can affect the growth of cosmic structures and the universe's evolution\cite{IDE2018,IDE2019,IDE1,IDE2}. We run simulations with different IDE parameters to study their impact on the large-scale structure and the galaxy distribution. These simulations are essential for testing the IDE model and for interpreting the CSST data in the context of alternative dark energy scenarios. 

The IDE model we have considered has been discussed in \cite{BM2015,MB2017,BM2017,Xiao2019}. In the Einstein frame, the scalar-tensor theory action of the model reads
\begin{equation}
\label{action}
\begin{split}
	\mathcal{S} =& \int d^4 x \sqrt{-g} \left[ \frac{M_{\rm Pl}^2}{2} R - \frac{1}{2} g^{\mu\nu}\partial_\mu \phi\, \partial_\nu \phi - V(\phi) + \mathcal{L}_{\rm SM}\right]\\
	&+ \int d^4 x \sqrt{-\tilde{g}} \mathcal{\tilde{L}}_{\rm DM}\left(\tilde g_{\mu\nu}, \psi\right)\,,
\end{split}    
\end{equation}
where the reduced Planck mass is $M_\text{Pl}=2.4\times 10^{18}$ GeV. The Lagrangian $\mathcal{L}_{\rm SM}$ represents terms from the standard model (SM), and dark energy is described by a quintessence scalar field $\phi$ with a potential $V(\phi)$, the last term $\mathcal{\tilde{L}}_{DM}$ corresponds to the dark matter sector which depends on the metric with the form
\begin{equation}
\label{DM metric}
\tilde g_{\mu\nu} = C(\phi) g_{\mu\nu} + D(\phi)\, \partial_\mu\phi\, \partial_\nu \phi\,  
\end{equation}
where $C(\phi)$ and $D(\phi)$ denote the conformal and disformal coupling factors, respectively. We can see dark matter and dark energy interaction clearly in the term $\mathcal{\tilde{L}}_{DM}$.  The interaction between DE and DM can be described by
\begin{equation}
\label{coupling function}
    Q=\frac{C_{,\phi}}{2C}T_{\rm DM}+\frac{D_{,\phi}}{2C}T_{\rm DM}^{\mu\nu}\nabla_\mu\phi\nabla_\nu\phi-\nabla_\mu\left[\frac{D}{C}T^{\mu\nu}_{\rm DM}\nabla_\nu\phi\right],\,
\end{equation}
where the subscript $(,\phi)$ denotes the derivative with respect to $\phi$. The pressureless DM has $T_{\rm DM}^{\mu\nu}$ to be its perfect fluid energy-momentum tensor, and $T_{\rm DM}$ is the corresponding trace. The interaction depends on the forms of $C(\phi), D(\phi), V(\phi)$. To be consistent with the discussions in \cite{BM2015,MB2017,BM2017,Xiao2019}, we choose exponential functional forms as
\begin{equation}
\label{coupling_choice}
C(\phi)=e^{2\alpha\kappa\phi},\;D(\phi)=D_m^4 e^{2\beta\kappa\phi},\;V(\phi)=V_0^4 e^{-\lambda\kappa\phi},   
\end{equation}
where $\alpha,\,D_m,\,\beta,\,V_0,$ and $\lambda$ are constants and $\kappa\equiv M_\text{Pl}^{-1}$.


The IDE simulations are performed using \textsc{ME-Gadget4}, an updated version of \textsc{ME-Gadget} implemented into \gadget4. \textsc{ME-Gadget} generalizes the \gadget code to work with non-standard cosmologies by introducing several phenomenological parameters in the hydrodynamical equations of structure formation, whose values can be derived for each given cosmological model. \rev{For the IDE models, the modifications of the cosmological evolution include changes to the Hubble parameter, the particle mass, the particle acceleration, and the gravitational constant. Detailed evolutions of these parameters in our simulated models are shown in~\ref{sec:IDE_evo}.} For each of the IDE simulations, \fof halos and \subfind subhaloes are identified using the \gadget4 pipeline at each snapshot. More detailed introductions of the algorithm are presented in \cite{IDE2018}. The ME-Gadget4 code is publicly available\footnote{https://gitee.com/shao-eor/me-gadget4}.

The simulations are listed in Table~\ref{tab:extension}. Conformal1 and Conformal2 represent the simulations that only include the $C(\phi)$ term, and $\alpha=0.03$ and $0.06$. Disformal1 and Disformal2 represent the simulations that only include $D(\phi)$ term, and $Dm=45$ and $105$, $\beta=0$. Quintessence1 and Quintessence2 represent the simulations that only include the $V(\phi)$ term, and $\lambda=0.8$ and $1.6$. The values are chosen following the convention and constraints from \cite{BM2017,Xiao2019}.


\subsection{Cosmic Neutrinos}
Neutrinos are standard model particles whose exact masses are still unknown. Neutrino oscillation experiments can only measure the square mass differences among the three neutrino species, while cosmological probes are at present the most sensitive way for constraining their total masses~(e.g.,\cite{EuclidNeutrino}). Neutrinos behave as radiation in the early universe, while massive neutrinos behave as dark matter at late time, leaving detectable signatures in both the cosmic microwave background (CMB) and the large scale structure. The most recent constraint from the DESI survey combined with CMB data reports an upper bound of $\sum m_\nu <0.072\mathrm{eV}$ at $95\%$ confidence level~\cite{DESI_cosmology}. 

On large scales, neutrino distributions are essentially degenerate with dark matter, and the detection of neutrino masses from galaxy surveys can only rely on their influences on intermediate to small scales where the free streaming of neutrinos induces a suppression on the non-linear matter clustering, necessitating the use of cosmological simulations. The large thermal velocities of neutrino particles, however, poses a significant challenge to including them in simulations (see \cite{AnguloHahn} for a recent review). To properly sample their initial velocities distribution in addition to the spatial distribution, a much larger number of $N$-body particles are needed in order to suppress shot noise. An alternative approach is to simulate the neutrinos as a fluid by solving the Boltzmann momentum equations on a grid, which is free from shot noise and can be more computationally efficient. This is the approach adopted in the neutrino simulations of the Jiutian extension runs, as implemented in the information optimized P$^3$M code \cube~\cite{cube}. The method decomposes the neutrino phase-space into three shells of constant speed to derive a closed set of fluid equations for neutrinos, taking into account the full nonlinear gravitational source terms~\cite{CubeNu}. We carry out simulations with two different neutrino masses, $\sum m_\nu=0.05$ and $0.1\mathrm{eV}$ and adopting two boxsizes of $1.5$ and $3\gpch$ using $2048^3$ dark matter particles, as listed in Table~\ref{tab:extension}. The neutrino density fields are also saved in the outputs, in addition to the dark matter particles. 

The Jiutian emulator runs also account for the existence of massive neutrinos, using a gauge transformation method which does not require the explicit inclusion of a neutrino component in the simulation. The neutrino runs in the extension module complement the emulator runs by also predicting the distribution of neutrinos explicitly, using an alternative method, and in larger boxes. \rev{Both methods can accurately reproduce the effect of neutrinos on the cold dark matter power spectrum. We refer interested readers to \cite{NeutrinoComp} for comprehensive comparisons of different neutrino simulation methods.}

\section{Summary and conclusions}\label{sec:summary}
The Jiutian simulations are a comprehensive suite of cosmological simulations conducted to support the development and scientific analysis of the CSST extra-galactic surveys. It consists of four complementary modules, designed to meet the science requirements of CSST surveys from different angles. The primary module meets the resolution and volume requirements of the surveys with three high resolution simulations in varying volumes. The emulator module targets the precision in cosmological parameter inference by carrying out hundreds of medium resolution simulations with different cosmological parameters and constructing a cosmological emulator. The reconstruction module offers detailed insights into the formation and evolution of galaxies in specific regions of the universe by carrying out zoom-in hydrodynamical simulations according to observationally constrained initial conditions. Lastly, the extension module explores major modifications to the \LCDM model, through simulations under alternative dark matter and dark energy models as well as different neutrino masses.

Diverse data products have been derived from the primary runs, including particle lightcones and snapshots, halo and subhalo catalogs, merger trees and galaxy catalogs, with broad prospective applications in many theoretical and observational studies. Significant developments have also been made in the corresponding post-processing pipelines. The \gaea SAM is further developed to work with subhalo catalogs and merger trees from the time domain subhalo finder \hbt, and a light-cone builder, \blic, is developed to produce subhalo and galaxy lightcones. The \lgalaxy SAM is improved to predict emission lines self-consistently in the galaxy spectrum, as well as to incorporate gravitational lensing effects induced by self-consistent density field in the galaxy lightcone. An extension to the SHAM technique is developed to match subhalos with observed galaxies in multiple properties, to produce an empirical galaxy lightcone catalog that can closely match the distribution of DESI galaxies. These developments yield four sets of independent galaxy lightcone catalogs.

Based on the \hbt subhalos in the primary runs, we verify the universality of the peak mass function of subhalos at various merger levels proposed in \cite{Jiang25}, down to a mass limit of $10^{-5}$ times the host mass. The results show that the level-1 subhao peak mass function (PMF) is consistent with being universal across host halo mass and redshift. The PMFs of high level subhalos decrease slightly towards lower redshifts, which can be modelled accurately with a redshift-dependent parameter accounting for the ratio between central mass and host mass at peak time.

 
The emulator runs sample eight cosmological parameters including the mass of neutrinos and the equation of state of dynamical dark energy. The wide and efficient coverage of parameter space and the uniform numerical specifications make the emulator runs particularly useful for comparative studies of structure formation under different cosmologies, and for the establishment of universal laws across cosmologies. The CSST emulator constructed from the emulator runs achieves 1\% precision in predicting the power spectrum for $k\leq 10 h\mathrm{Mpc}^-1$, out-performing the Mira-Titan and Euclid emulators. 

Based on constrained initial conditions from the SDSS surveys, the reconstruction runs are able to match not only the galaxy distributions but also the gas properties in specific regions of the observed universe such as the Coma cluster. With the implementation of different subgrid models in different runs, these simulations provide a direct testbed for galaxy formation models while segregating environmental effects and cosmic variance.

The extension runs allow for the study of structure formation in more general cosmological models, including WDM, $f(R)$ gravity, interacting dark energy and massive neutrino models. These runs also pose unique challenges to the simulation and post-processing techniques. 
Special techniques have been implemented or developed in the Jiutian simulations to cope with these challenges. The WDM simulations have already been used to generalize analytical subhalo distribution models from CDM to WDM cosmologies.

The Jiutian simulations have already been used in a number of studies on cosmology and galaxy formation, as well as in the development of the CSST data analysis pipelines. These simulations, loosely organized under the Jiutian project, form a collaborative and open simulation library for the community. The total data amounts to 8PB. High level data products will be gradually released at \url{https://jiutian.sjtu.edu.cn}, while low level data will be accessible through direct communications. Anyone interested in using the simulations can get in touch with one of the collaboration members to be redirected to the corresponding resources. 

\rev{One important avenue that Jiutian has not explored is the generation of covariance matrices for largescale structure statistics, which is best computed from a large number of independent simulation realizations under the same cosmology (e.g., \cite{AbacusRun}). Due to the high computational cost of full $N$-body simulations, many previous studies, however, rely on semi-analytical or semi-empirical calculations (e.g., \cite{DESIcov, EUCLIDcov}) based on approximate simulations (e.g., FastPM~\cite{FastPM}, PINOCCHIO~\cite{PINNOCHIO}, EZmock~\cite{EZmock}, inverse-Gaussianization~\cite{InverseGaussian}). In this context, the large number of Jiutian simulations with different cosmologies can provide a high-fidelity benchmark sample to calibrate the statistics from fast methods. The large box runs (e.g., Jiutian-2G) could also be used for covariance calculations on small to intermediate scales.}


\Acknowledgements{This work was supported by China Manned Space Project (No.\ CMS-CSST-2021-A03) and National Key R\&D Program of China (2023YFA1607800, 2023YFA1607801). YSQ acknowledges support from Key R\&D Program of Zhejiang, China (Grant No. 2024SSYS0012).}

\InterestConflict{The authors declare that they have no conflict of interest.}

\bibliographystyle{scpma}
\bibliography{jiutian}{}

\begin{appendix}
\section{Cosmological evolution in the IDE models}\label{sec:IDE_evo}
\begin{figure*}
    \centering
    \includegraphics[width=\textwidth]{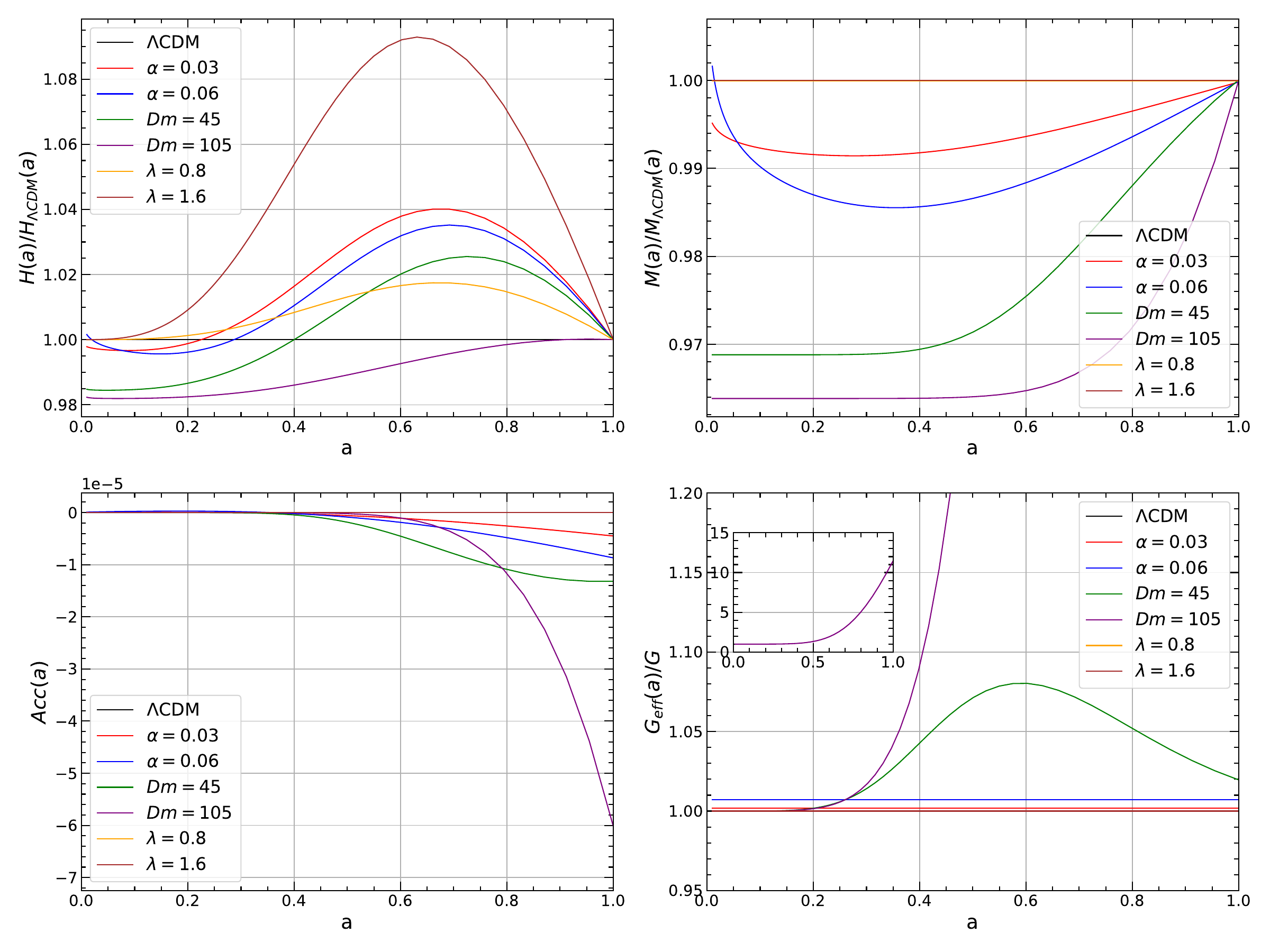}
    \caption{\rev{The evolutions of Hubble parameter (top left), particle mass (top right), additional acceleration (bottom left), and effective gravitational constant (bottom right) as functions of scale factor, compared to that in the $\Lambda$CDM model. Note that in the bottom right panel, the $Dm=105$ case is illustrated in a separate small window due to its larger dynamic range. 
    }}
    \label{fig:IDE-evo}
\end{figure*}

\rev{In the simulations of the IDE models, the Hubble parameter is changed due to modification to the Friedmann equation, the particle mass is changed due to the energy transfer between dark matter and dark energy, an extra velocity-dependent particle acceleration is induced due to the momentum transfer between dark matter and dark energy, and the effective gravitational constant deviate from the Newtonian gravitational constant due to the perturbation in dark energy. These values are all evolving as a function of the scale factor. In Fig.~\ref{fig:IDE-evo} we show these evolutions in the Jiutian IDE simulations, as used in the \textsc{ME-Gadget4} code. Their detailed meanings and some tests on the calculations can be found in \cite{IDE2018,BM2017,Xiao2019}.}

\end{appendix}

\end{multicols}
\end{document}